\documentclass[10pt,
journal=jctcce, % jctc
manuscript=article,
layout=twocolumn
]{achemso}

\usepackage[version=3]{mhchem} % Formula subscripts using \ce{}
\usepackage{physics}
\usepackage{multirow}

\newcommand{\matr}[1]{\mathbf{#1}}
\usepackage{graphicx}
\usepackage{amsmath,bm}
\usepackage{upgreek}
\graphicspath{ {figures/} }

\altaffiliation{4400 Fifth Ave., Pittsburgh PA 15213, USA}
\author{Haichen Li}
\author{Christopher Collins}
\author{Matteus Tanha}
\affiliation[chem.cmu]{Department of Chemistry, Carnegie Mellon University, Pittsburgh, PA 15213}
\author{Geoffrey J. Gordon}
\affiliation[ml.cmu]{Machine Learning Department, Carnegie Mellon University, Pittsburgh, PA 15213}
\author{David J. Yaron}
\email{yaron@cmu.edu}
\affiliation[chem.cmu]{Department of Chemistry, Carnegie Mellon University, Pittsburgh, PA 15213}

\title[DFTB layer for deep learning]
{A Density Functional Tight Binding Layer for Deep Learning of Chemical Hamiltonians}

\begin{document}

\begin{abstract}
Current neural networks for predictions of molecular properties use quantum chemistry only as a source of training data. This paper explores models that use quantum chemistry as an integral part of the prediction process. This is done by implementing self-consistent-charge Density-Functional-Tight-Binding (DFTB) theory as a layer for use in deep learning models. The DFTB layer takes, as input, Hamiltonian matrix elements generated from earlier layers and produces, as output, electronic properties from self-consistent field solutions of the corresponding DFTB Hamiltonian. Backpropagation enables efficient training of the model to target electronic properties. Two types of input to the DFTB layer are explored, splines and feed-forward neural networks. Because overfitting can cause models trained on smaller molecules to perform poorly on larger molecules, regularizations are applied that penalize non-monotonic behavior and deviation of the Hamiltonian matrix elements from those of the published DFTB model used to initialize the model. The approach is evaluated on 15,700 hydrocarbons by comparing the root mean square error in energy and dipole moment, on test molecules with 8 heavy atoms, to the error from the initial DFTB model.  When trained on molecules with up to 7 heavy atoms, the spline model reduces the test error in energy by 60\% and in dipole moments by 42\%. The neural network model performs somewhat better, with error reductions of 67\% and 59\% respectively. Training on molecules with up to 4 heavy atoms reduces performance, with both the spline and neural net models reducing the test error in energy by about 53\% and in dipole by about 25\%.
\end{abstract}

%Model performance is evaluated by training on molecules with up to either 4 or 7 heavy atoms and evaluating the root mean square error on molecules with 8 heavy atoms.

\section{Introduction}\label{sec:intro}
Machine learning (ML) has the potential to predict molecular properties at low computational cost, making it possible to rapidly search chemical space for optimal systems.\cite{von2017machine,guzikInverse2018,rampi2015,ex_chem_screen2,genetic} A powerful strategy for developing ML models begins by using quantum chemistry (QC) to generate a set of data that spans the chemical space of interest.\cite{rupp2012fast,gdb17,clean_energy2} The ML model is then trained to reproduce the QC data with acceptable accuracy and with substantially reduced computational cost.\cite{atomization_review} The ML models explored to date use, as input, either features derived solely from the structure of the molecular system\cite{rupp2012fast,atomization_review,split_energies,nn_exp,local_coulomb,yao2017intrinsic,systematic,collins2018constant}, or, in the case of $\Delta$-machine-learning models,\cite{early_delta,ediz_yaron,ramakrishnan2015big,delta_ml_electronic} a combination of structural features and results from lower-cost quantum chemical methods. While quantum mechanics is used to generate the training data, the form of the ML model does not itself incorporate aspects of quantum mechanics. In the models explored here, quantum chemistry is an integral part of the prediction process. The approach may be viewed as an extension of semiempiricial QC in which ML models are used to generate the empirical parameters for the model Hamiltonian.

The models explored here are developed and tested on a subset of the ANI-1 data set\cite{smith2017anidata}, which includes small organic molecules distorted from their equilibrium positions. The ANI-1 potential\cite{smith2017ani} is a neural network that achieves high accuracy on the entire ANI-1 data set. The network architecture is similar to that introduced by Behler and Parrinello\cite{behler2007generalized} and that obtains high performance on a range of chemical systems.\cite{behler2015tutorial,boes2016neural,shen2016multiscale,artrith2016implementation} Incorporating QC into the neural network may be viewed as a way to incorporate domain knowledge into the ML model. Semiempirical QC models are able to describe a broad range of chemical phenomena, including valency, bond formation, and aromaticity. Semiempirical QC methods have also had considerable quantitative success.\cite{dewar1985development,stewart2013,thiel2014semiempirical,repasky2002pddg,dftbReview2014} Incorporating this domain knowledge into the neural network may help lower the amount of data needed to train the model and help improve transfer between systems. However, we note that in domains such as computer vision and natural language processing, this domain knowledge has been pruned from the ML models as the amount of training data and sophistication of the ML approaches have increased.\cite{girshick2015fast,ren2015faster,redmon2016you,collobert2008unified,collobert2011natural}

%This model form is within the large family of semiempirical models\cite{wu2017self} that have been under development for over four decades.\cite{pople1965approximate,pople1967approximate,dewar1977ground,zerner1991semiempirical,thiel1996perspectives,stewart2008application,stewart2009application,foster2010empirically,cui2014density}
The QC portion of the model developed here is based on self-consistent-charge Density-Functional Tight-Binding (DFTB) theory.\cite{elstner1998self,koskinen2009density,nishizawa2016three} DFTB uses a minimal-basis, valence-electron only Hamiltonian, with atomic point charges used to describe Coulomb interactions between atoms. A unique characteristic of DFTB is that the parameters for the electronic Hamiltonian are obtained through a non-empirical approach.\cite{sattelmeyer2006comparison}  This approach begins by generating QC solutions for isolated atoms placed in a simple electrostatic potential that constrains the radial spread of the electron distribution. The resulting atomic wavefunctions are then used to derive parameters for the electronic Hamiltonian.\cite{yang2008description,dolgonos2009improved,wahiduzzaman2013dftb,oliveira2015dftb,zheng2007parameter}  DFTB typically also includes a repulsive potential that is handled empirically by fitting to QC or experimental results for molecules or periodic systems.\cite{knaup2007initial,gaus2009automatized,bodrog2011automated,lourencco2016fasp}

Below, we implement DFTB as a layer for deep learning, using the TensorFlow deep learning framework.\cite{abadi2016tensorflow} The DFTB layer takes, as input, values for the Hamiltonian matrix elements that define the DFTB model and generates, as output, molecular properties that are self-consistent-field solutions of the corresponding DFTB Hamiltonian. The DFTB layer supports backpropagation, allowing the model parameters to be updated efficiently. Below, training on 12,400 small organic molecules, sampled from the ANI-1 data set\cite{smith2017anidata}, takes about 5 hours on 6 processor cores\cite{bridges}.

Our focus here is on the construction and characterization of the DFTB layer itself, which is agnostic with respect to the form of the layers used to generate the DFTB matrix elements. To help characterize the DFTB layer, we consider two different types of earlier layers. In the spline model, we use spline functions to allow the matrix elements to be functions of the inter-atomic distance, $r$. In the neural network model, we use feed-forward neural networks (FFNN) to allow the matrix elements to be more general functions of the molecular geometry. We will refer to the combination of the input layers and the DFTB layers as a DFTB-ML model.

Because DFTB-ML models are highly flexible, we begin by initializing the input layers to the matrix elements from a published DFTB parameterization.\cite{elstner1998self} Training of the DFTB-ML model may be viewed as refining this initial DFTB model. Although performance on the molecules in the training set improves as training progresses, performance on test molecules may begin to degrade with continued training. Below, such overfitting is especially prevalent when the model is trained on smaller molecules and then applied to larger molecules. We consider two different regularizations to reduce such overfitting. The first applies only in the spline model and constrains the Hamiltonian matrix elements to have a monotonic dependence on $r$. The second regularization, which is much more effective and applies to both spline and FFNN models, penalizes deviation from the initial DFTB parameters.  This penalty is a regularization that limits the model flexibility and helps in the transfer of models from smaller to larger molecules.

One motivation for considering transfer from smaller to larger molecules is that this relates to a potential advantage of building quantum mechanics directly into the ML model form. During training, we modify only matrix elements that describe short range interactions between atoms. The longer-range interactions are described via Coulomb's law. This transition from empirical interactions at short range to physics-based interactions at long range may improve the degree to which models trained on small molecules are able to transfer to larger systems.

%The benefits and risks of this approach to ML model development are explored below using a subset of the molecules in the ANI-1 data set.\cite{smith2017anidata} ANI-1 was used to train deep learning models for distortion energies of organic molecules.\cite{smith2017ani} The subset used here includes molecules with up to 8 heavy atoms, consisting of only the elements H, C and O. The QC data was regenerated so we could obtain the atomic charges against which to train the ML model.

\section{Related work}
In current DFTB models, empirical fits are typically limited to the repulsive potential, a classical potential whose energy is added to the energy obtained from the DFTB electronic Hamiltonian. A number of approaches have been developed to help automate fits of the DFTB repulsive potential to energies and forces obtained from {\it ab initio} QC.\cite{knaup2007initial,gaus2009automatized,bodrog2011automated,lourencco2016fasp} The repulsive potential is typically written as a sum of interatomic potentials that are nonzero over a limited range, typically just beyond the range of a covalent bond. The functional forms are also often restricted, for example, to sums of exponentials. More recently, unsupervised learning has been used to develop repulsive potentials with more general model forms.\cite{kranz2018generalized} Because our focus here is on the electronic Hamiltonian, we use spline functions to obtain a flexible, but relatively simple, form for the repulsive potential.

A number of recent DFTB parameterizations have, in addition to fitting the repulsive potential, empirically adjusted parameters that define the electronic Hamiltonian during the fitting process\cite{chou2015automatized,cawkwell2017,balintDFTBauto2018,YueDFTBauto2018,huran2018efficient}. Adjusted parameters include those that define the constraining potential and electron density cutoffs in the standard approach utilized to construct the DFTB electronic Hamiltonian from QC solutions for isolated atoms \cite{chou2015automatized,balintDFTBauto2018,YueDFTBauto2018}. Empirical fits have also adjusted the atomic orbital energies, and the Hubbard parameters that specify electron-electron repulsion\cite{chou2015automatized,cawkwell2017,balintDFTBauto2018,YueDFTBauto2018}. Matrix elements between atoms have also be adjusted by fitting analytic forms that describe the dependence of these matrix elements on $r$ and that involve 2\cite{cawkwell2017} or between 12 and 15\cite{huran2018efficient} free parameters per matrix element type. This past work suggests that adjusting the electronic Hamiltonian can lead to significant improvements in model accuracy.

Empirical fits of the electronic Hamiltonian have used a number of optimization schemes. In a dual loop approach, optimization of the electronic Hamiltonian alternates with optimization of the repulsive potential, using different objective functions for each of these two optimization loops\cite{balintDFTBauto2018,YueDFTBauto2018}. Simultaneous optimization of all fitting parameters has also been done using gradient free optimization methods such as swarm optimization\cite{chou2015automatized}, simulated annealing\cite{cawkwell2017}, and pattern search\cite{huran2018efficient}. These optimization methods do not require the gradients of the molecular properties with respect to model parameters to be computed. This has the advantage of allowing parameters to be adjusted to fairly complex targets such as minimum-energy structures, lattice parameters, and energy differences between polymorphs. The approach developed here uses back propagation to efficiently compute gradients. This allows flexible models, that involve a large number of parameters, to be trained on large sets of molecular data. However, the targets are limited to molecular properties such as energy and dipole that are functions only of the input molecular geometries.

An alternative, ML-based approach, for optimizing parameters in the electronic Hamiltonian has been applied to the OM2 semiempirical QC model.\cite{weber2000orthogonalization,dral2015machine} This iterative approach considers one parameter at time. For each molecule in the data set, the value of the parameter that minimizes the error for that individual molecule is obtained. Kernel Ridge Regression (KRR) is then used to predict this optimal value, using only the molecular structure. This approach reduced mean absolute errors in atomization energies on test molecules from 6.7~to~1.3~kcal/mol.

\section{Methods}

\subsection{DFTB Hamiltonian}\label{sec:dftbH}

The DFTB model originates from approximations applied to the Kohn-Sham equations of density functional theory.\cite{elstner1998self}  Here, we use the model as an empirical form for model fitting and so describe the model from an empirical perspective.  The single-electron wavefunctions or molecular orbitals, $\Psi_a$, are expressed in a minimal atomic basis, $\phi_i$,
\begin{equation}\label{eq:orbs}
   \Psi_a = \sum_i^{N_{basis}} C_{i,a} \phi_i ,
\end{equation}
where the $N_{basis}$ atomic basis functions include only valence orbitals, {\it e.g. 1s} on H and {\it 2s, 2p} on second-row elements.

The Kohn-Sham equations may be written as the following eigensystem,
\begin{equation}\label{eq:KSequations}
   \sum_j^{N_{basis}} \left[ H_{i,j} + H^{(2)}_{i,j}(\Delta q) - \epsilon_a S_{i,j} \right] C_{j,a} = 0, \forall a,i
\end{equation}
where $H_{i,j}$ and $S_{i,j}$ are matrix elements of the one-electron Hamiltonian operator and overlap operator, respectively, between atomic orbitals $i$ and $j$, and $\epsilon_a$ are the Kohn-Sham orbital energies sorted from low to high.  $H^{(2)}_{i,j}(\Delta q)$ describes the two-electron interactions as interactions between charge fluctuations in the atomic shells, $\Delta q$,
\begin{equation}\label{eq:H2DFTB}
\footnotesize
  H^{(2)}_{i,j}(\Delta q) = \frac{1}{2} S_{i,j}\sum_{\alpha}^{N_{shells}}
   \left( \gamma_{shell\left(i\right),\alpha} + \gamma_{shell\left(j\right),\alpha} \right) \Delta q_{\alpha}
\end{equation}
where $\alpha$ indexes over the $N_{shells}$ atomic shells ({\it e.g. 1s, 2s, 2p}) in the molecule,  $shell\left(i\right)$ is the shell of atomic orbital $i$, $\Delta q_{\alpha}$ is the charge fluctuation of shell $\alpha$, and $\gamma_{\beta,\alpha}$ is the Coulomb interaction between shells $\beta$ and $\alpha$.

The charge fluctuations are obtained from the Mulliken population of each atomic shell. These may be obtained from the electronic density matrix,
\begin{equation}
    \rho_{i,j} = \sum_a n_a C_{i,a} C_{j,a}
\end{equation}
where $i$ and $j$ index atomic orbitals and $n_a$ is the occupation of the $a^{th}$ molecular orbital. The Mulliken charge of each atomic shell is then
\begin{equation}\label{eq:qshell}
  q_{\alpha} = - \sum_{i \in shell\left(\alpha\right) } \sum_{j}
      \rho_{i,j} S_{i,j}
\end{equation}
The charge fluctuation of the $\alpha^{th}$ shell is then,
\begin{equation}\label{eq:dqshell}
 \Delta q_{\alpha} =  q_\alpha - q^{(0)}_\alpha
\end{equation}
where $q^{(0)}_\alpha$ is the charge of atomic shell $\alpha$ in the isolated, neutral, atom. The values of $q^{(0)}_\alpha$ are constants that may be taken as part of the model parameterization.

In Section~\ref{sec:tensors} below, we convert $H^{(2)}(\Delta q)$ from the summation over atomic shells in Eq.~\ref{eq:H2DFTB} to a summation over atomic orbitals. With each atomic orbital $\phi_i$, we introduce a charge
\begin{equation}\label{eq:qatom}
  Q_{i} = - \sum_{j} \rho_{i,j} S_{i,j}
\end{equation}
and a charge fluctuation
\begin{equation}\label{eq:dqatom}
 \Delta Q_{i} =  Q_i - Q^{(0)}_i
\end{equation}
where the reference charge for the atomic orbitals, $Q^{(0)}$, is obtained by distributing the reference charge for the shells, $q^{(0)}$ of Eq.~\ref{eq:dqshell}, equally across all atomic orbitals in the respective shell. $H^{(2)}_{i,j}(\Delta q)$ of Eq.~\ref{eq:H2DFTB} may then be written
\begin{equation}\label{eq:H2DFTB2}
  H^{(2)}_{i,j}(\Delta q) = \frac{1}{2} S_{i,j}\sum_{k} \left( G_{i,k} + G_{j,k} \right) \Delta Q_{k}
\end{equation}
with
\begin{equation}\label{eq:Gmatrix}
    {G}_{i,j} = \gamma_{shell\left (i \right), shell\left( j \right)}
\end{equation}

The total electronic energy is then
\begin{equation}\label{eq:Eelec}
    E_{elec} = \sum_{i,j} \rho_{i,j} H_{i,j} +
     \frac{1}{2} \sum_{i,j} \Delta Q_{i} G_{i,j} \Delta Q_{j}
\end{equation}

The total energy of the system also includes a classical potential energy term referred to as the repulsive potential because it is intended to include the repulsive interaction between the atomic cores that are not included in the electronic Hamiltonian
\begin{equation}\label{eq:dftbErep}
    E_{rep} = \sum_{A>B} R_{Z_A,Z_B}\left( \left| \matr{r}_A - \matr{r}_B \right| \right)
\end{equation}
where $A$ and $B$ label atoms, $r_A$ is the cartesian position of the $A^{th}$ atom, and $R_{Z_A,Z_B}$ is a function that depends on the elements, indicated by atomic numbers $Z_A$, of the atoms.

We also include a reference energy such that the total energy is given by,
\begin{equation}\label{eq:Eref}
    E_{tot} = E_{elec} + E_{rep} + E_{ref}.
\end{equation}
DFTB parameterizations typical consider only the change in energy due to geometric distortion.\cite{lourencco2016fasp} This can be done, for example, by fitting to the energy differences between molecular geometries or by fitting to atomic forces. Fitting to distortion energies may be viewed as assigning a separate reference energy to each isomer under consideration. Below, we instead use a reference energy based on atom counts,
\begin{equation}\label{eq:Erefdef}
    E_{ref} = \sum_{Z=1,6,8}{p_{Z} N_{Z}} + p_c
\end{equation}
where the sum is over the elements (H,C,O), $p_Z$ is a parameter associated with each element, $N_Z$ is the number of occurrences of that element in the molecule, and $p_c$ is a single parameter that sets the overall zero of energy. Eq.~\ref{eq:Erefdef} has the advantage of allowing the model to predict energy differences between any molecules composed of elements present in the training data.

\subsection{Matrix element models}\label{sec:models}

The parameters of a DFTB model are used to construct the matrices $\matr{H}$, $\matr{S}$ and $\matr{G}$ of Eqs.~\ref{eq:KSequations} and~\ref{eq:H2DFTB2}, and the repulsive potentials, $R_{Z_A,Z_B}$ of Eq.~\ref{eq:dftbErep}. In practice, a DFTB parameterization is specified through a set of files that list constants for the on-atom terms and provide the between-atom terms as either a parametric or tabulated function of the interatomic separation, $r$. Our goal is to provide an efficient means to learn more flexible forms for these parameters. Below, we use either splines or feed-forward neural networks as examples of such flexible forms. However, our intent is to support any model form that can be implemented efficiently in a deep-learning framework such as Tensorflow. We therefore divide the responsibilities between ``models'' that generate information of the type currently extracted from the files that define a DFTB parameterization, and a ``DFTB layer'' that uses this information to predict molecular properties. The models are implemented as layers that provide input to the DFTB layer. The DFTB layer is agnostic with regards to the form of these previous layers, requiring only that these earlier layers produce the information in the order specified in Section~\ref{sec:tensors}.

Although our implementation allows deep learning to be used to construct $\matr{S}$, we currently fix $\matr{S}$ to values from an existing DFTB parameterization\cite{elstner1998self} and so do not explore the potential benefits and challenges associated with empirically fitting $\matr{S}$. Also, because our focus is on the electronic Hamiltonian, we restrict the repulsive potentials, $R_{Z_A,Z_B}$ of Eq.~\ref{eq:dftbErep}, to one-dimensional spline functions.

Table~\ref{tab:models} lists the models needed to construct the matrices $\matr{H}$ and $\matr{G}$ for molecules composed of the elements H, C and O. These matrices are assembled from ``blocks'', with diagonal blocks being between orbitals on a single atom and off-diagonal blocks being between orbitals on two different atoms. For diagonal blocks of $\matr{H}$, the models generate the energy of the $s$ and $p$ atomic orbitals. We allow these orbital energies to depend on the environment of the atom,\cite{schutt2018machine} but assume that all orbitals of a given shell on a given atom have the same energy. For off-diagonal blocks of $\matr{H}$, the models generate the matrix elements between atomic orbitals that are aligned along the axis connecting the two atoms. We will refer to these as ``aligned'' matrix elements. For first-row elements, the unique aligned elements are $ss$, $sp$, $pp_\sigma$, and $pp_\pi$. These aligned matrix elements may depend on the environments of the associated atoms. Slater-Koster (SK) rotations are linear transformations that rotate these aligned elements into blocks of $\matr{H}$ between atoms. The DFTB layer described below carries out the SK rotations, with the input layers generating only the aligned matrix elements. For both diagonal and off-diagonal blocks, the models for $\matr{G}$ generate the Coulomb interaction, $\gamma$ of Eq.~\ref{eq:H2DFTB}, between the various shells, $ss$, $sp$ and $pp$. Because Coulomb interactions between atoms in DFTB are between point charges associated with each atomic shell, Eq.~\ref{eq:H2DFTB}, SK rotations are not necessary for $\matr{G}$.

%Our goal is to use deep-learning to allow the DFTB parameters to be more general functions of the molecular structure. To do this, we implement DFTB as a layer for deep learning that takes, as inputs, predictions from ``models'' that generate the diagonal matrix elements and the aligned off-diagonal matrix elements. Table~\ref{tab:models} lists the models needed for molecules containing the element C,H and O.  We separately consider models that generate diagonal (on-atom) versus off-diagonal (between atoms) matrix elements, as these models will typically use different sets of features and have different architectures. Also, the overlap operator, $\matr{S}$, is not included in Table~\ref{tab:models} because we currently fix $\matr{S}$ to values from an existing DFTB parameterization\cite{elstner1998self}. In DFTB, the diagonal models would generate a constant value that depends only on element and orbital type. Below, we optionally allow them to depend on the environment of each individual atom. In DFTB, off-diagonal models are functions of only the separation between the atoms, $r$. Below, we optionally allow models to take the molecular context into account.

\begin{table*}[t]
\centering
\caption{Types of aligned DFTB matrix elements. Block type refers to blocks of the operator on a single atom (diagonal) or between two atoms (off-diagonal). Orbital type refers to the orbitals involved in the aligned matrix elements. Blocks of the operators on or between atoms can be generated from these aligned matrix elements through Slater-Koster rotations. The last column lists the number of models of each type needed for molecules containing the elements C,H, and O.}
\label{tab:models}
\begin{tabular}{lllll}
\hline\hline
operator        & block type   & elements     & shell type                   & models for C,H,O \\
\hline
$\matr{H}$      & diagonal     & $Z$          & $s$, $p$                          & 5                \\
$\matr{H}$      & off-diagonal & $Z_1$, $Z_2$ & $ss$, $sp$, $pp_\sigma$, $pp_\pi$ & 18               \\
$\matr{G}$ & diagonal     & $Z$          & $ss$, $pp$, $sp$                  & 7                \\
$\matr{G}$ & off-diagonal & $Z_1$,$Z_2$  & $ss$, $pp$, $sp$                  & 15    \\
\hline
\end{tabular}
\end{table*}

%This includes the following:
%\begin{description}
%\item[Atomic orbital energies ($E^{atomic}_{\alpha}$)] Diagonal matrix elements of the one-electron Hamiltonian operator, $H_{\mu,\mu}$ for orbitals $\mu$ belonging to atomic shell $\alpha$. These are taken as constant in DFTB.
%\item[One electron Hamiltonian between atomsenergies ($E^{atomic}_{\alpha}$)] Diagonal matrix elements of the one-electron Hamiltonian operator, $H_{\mu,\mu}$ for orbitals $\mu$ belonging to atomic shell $\alpha$. These are taken as constant in DFTB.
%\end{description}

%Creating a DFTB layer for a deep learning network brings up three main challenges.
%\begin{enumerate}
%\item Implementing DFTB as a set of tensor operations, such that TensorFlow or other deep learning platform may be used to evaluate the gradients needed o efficiently train the parameters via back propagation.
%\item Handling the self-consistent-field (SCF) character of the DFTB solutions while training the network.
%\item Addressing numerical instabilities that may arise due to back propagation through the eigensystem of Eq.~\ref{eq:KSequations}, or instabilities due to nonconvergence of the SCF iterative solution process.
%\end{enumerate}
%These challenges are addressed in each of the following three subsections.

\subsection{Expressing DFTB as tensor operations}\label{sec:tensors}
Training of the DFTB model requires a large number of gradient descent updates to the model parameters. To make these updates efficient, we implement DFTB as a series of tensor operations in the Tensorflow deep learning framework\cite{abadi2016tensorflow}. Each update to the parameters is based on a set of molecules that we will refer to as a minibatch. In Tensorflow, a computational graph is first constructed that specifies the series of tensor operations required to make predictions for a single minibatch. Data that depends on the particular molecules in the minibatch is then fed into this graph. In a forward pass through the graph, the molecular properties are predicted based on the current model parameters, and the difference between predicted and target properties is used to compute a loss. In a backward pass, the gradient of the loss with respect to model parameters is computed and used to update the model parameters. Here, we use Tensorflow's ADAM optimizer\cite{kinga2015method} to update the parameters based on the computed gradients, with a learning rate of $10^{-5}$, first moment exponential decay rate of $\beta_1 = 0.9$, second moment exponential decay rate of $\beta_2 = 0.999$, and numerical stability constant $\epsilon = 10^{-8}$.

An epoch of training corresponds to a forward and backward pass performed on each minibatch in the training data. The form of the computational graph depends only on the empirical formulas of the molecules included in the minibatch. To allow us to use a single graph during model training, the molecules in each minibatch have identical empirical formulas, and the atoms in each molecule are sorted to have identical orders of elements. (This restriction on the empirical formulas of the minibatches could be relaxed by sharing model parameters between multiple computational graphs.)

%We refer to this implementation of DFTB as a layer for deep learning because it takes model parameters from previous layers as input (Figure~\ref{fig:tfgraph}a). The DFTB layer is agnostic with regards to the form of these previous layers, requiring only that the layers produce the aligned DFTB matrix elements of Table~\ref{tab:models} in a specified order.

As discussed in Section~\ref{sec:models}, the input to the DFTB layer comes from models that produce the aligned matrix elements of Table~\ref{tab:models} in a specific order. This order is specified during construction of the computational graph and depends only on the sequence of empirical formulas in a minibatch. The specification consists of a list of tuples that specify the type of aligned matrix element (as in Table~\ref{tab:models}), the index of the molecule within the minibatch and the indices of the respective atoms within that molecule. The layers that provide input to the DFTB layer may use these specifications to implement a wide variety of models. The list is ordered by type of aligned matrix element so that the input layers may produce all matrix elements of a given type with a single set of tensor operations. Consider, for example, the use of a separate feed-forward neural network for each of the matrix elements types, or models, in Table~\ref{tab:models}. From the molecular geometries in a particular minibatch, a list of feature vectors may be created that, when fed into the neural network for a model, produces the aligned matrix elements in the specified order. For diagonal elements, the feature vectors may describe the environment of the atom associated with that matrix element. For off-diagonal elements, the feature vectors may describe the environment of the pair of atoms associated with that matrix element.

\begin{figure}[ht]
\centering
\includegraphics[width=0.95\columnwidth]{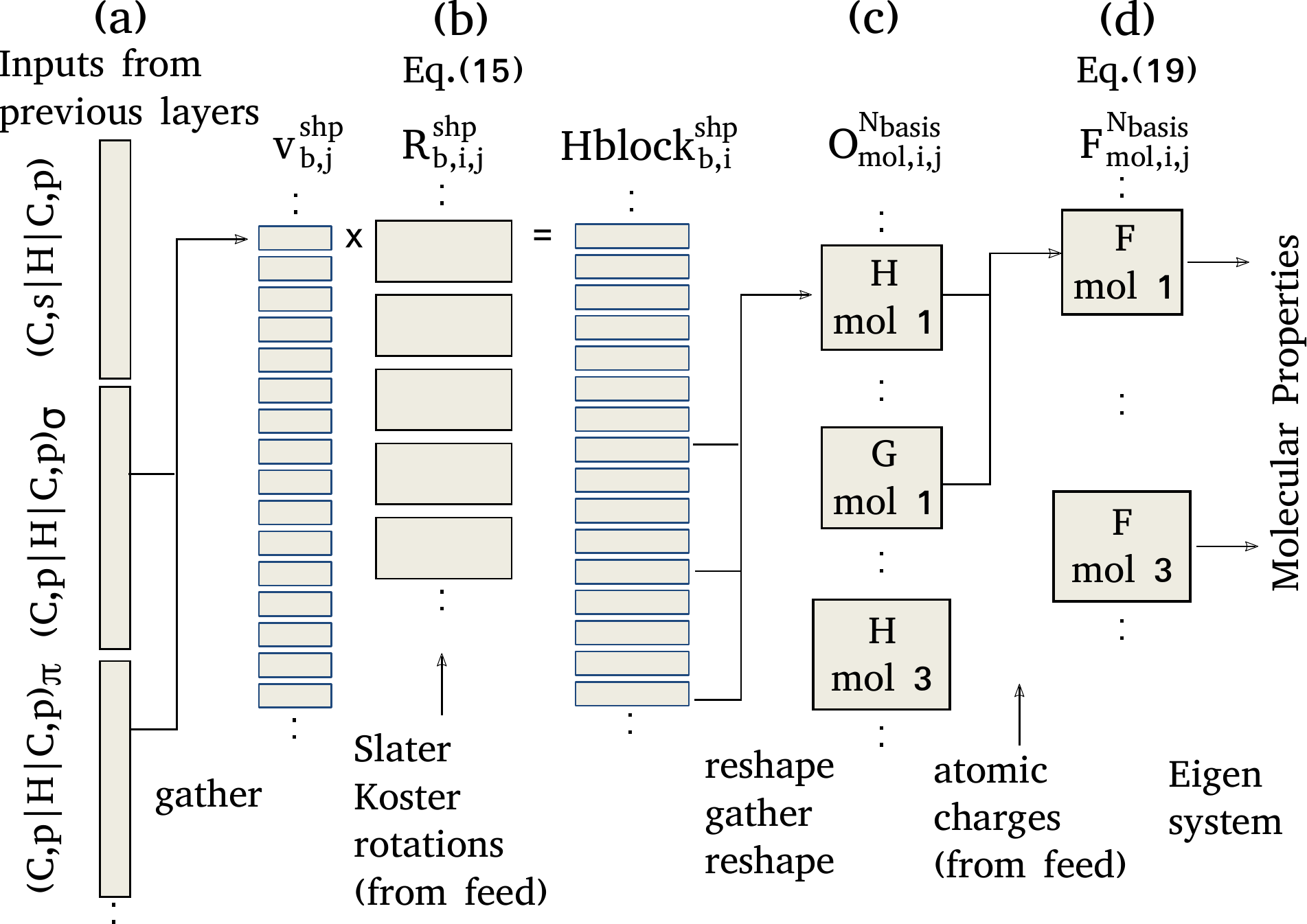}
\caption{Schematic representation of the DFTB layer. (a) The aligned matrix elements are input from previous layers, ordered as specified during construction of the computational graph. (b) Tensorflow gather and reshape operations are used to rearrange the input matrix elements and to carry out Slater-Koster rotations. (c) The resulting operator blocks are assembled into operators for each molecule. (d) Fock operators are constructed using atomic charges fed into the graph and molecular properties are predicted and output from the layer.}
\label{fig:tfgraph}
\end{figure}

The initial operations in the DFTB layer carry out the Slater-Koster (SK) rotations (Figure~\ref{fig:tfgraph}b). The rotations are written as batch matrix multiplies, which can be handled efficiently in Tensorflow or other deep-learning framework,
\begin{equation}\label{eq:SKRotation}
    {\bf Hblock}^{shp}_{b,i} = \sum_{j} {\bf R}^{shp}_{b,i,j} \matr{v}^{shp}_{b,j}
\end{equation}
The superscript $shp$ refers to the shape of the individual SK rotation matrices, i.e., the dimensions of indices $i$ and $j$ in Eq.~\ref{eq:SKRotation}. For example, $shp=3 \times 1$ arises when rotating the single aligned element of orbital type $sp$ to the three matrix elements in the operator block between $s$ and $p$ orbitals. $shp=9 \times 2$ arises when rotating the two aligned elements $pp_\sigma$ and $pp_\pi$ to the $3 \times 3$ operator block between $p$ orbitals. The index $b$ in Eq.~\ref{eq:SKRotation} labels blocks of the $H$ operator. For computational efficiency, a single evaluation of Eq.~\ref{eq:SKRotation} carries out all required SK rotations of a given shape, i.e., $b$ in Eq.~\ref{eq:SKRotation} runs over all blocks in the minibatch requiring SK rotations with shape $shp$. This requires substantial rearrangements of the values input to the DFTB layer. These are done using Tensorflow's {\it gather} and {\it reshape} operators. Given a vector $\matr{x}$ and a list of integers $\matr{I}$, {\it gather} returns $y_j$ = $x_{I_j}$. {\it Reshape} transforms a $N$-dimensional tensor into a flattened $1$-dimensional form and vice versa. For each shape, $shp$ in Eq.~\ref{eq:SKRotation},  a flattened view of $v^{shp}_{b,j}$ is gathered from the input and reshaped into a 2-dimensional tensor. The batch matrix multiply of Eq.~\ref{eq:SKRotation} then generates flattened views of the operator blocks, $\matr{Hblock}^{shp}_{b,i}$, ordered by a block index $b$.

An analogous approach is used to assemble blocks of $\matr{G}$. However, because SK rotations are not needed for $\matr{G}$, the values are only gathered and reshaped into $\matr{Gblock}^{shp}_{b,i}$.

The next operations in the DFTB layer assemble the operator blocks into matrices for the operators $\matr{H}$ and $\matr{G}$ of each molecule in the minibatch (Figure~\ref{fig:tfgraph}c). For computational efficiency, all operators of a given dimension, $N_{basis}$, are assembled through a single gather operation. The operator blocks, ${\bf Hblock}^{shp}_{b,i}$ of Eq.~\ref{eq:SKRotation}, are first flattened and concatenated into a single vector that holds results for all shapes, $shp$. For each $N_{basis}$, a flattened version of all operators with dimension $N_{basis}$ are then gathered and reshaped into $\matr{O}^{N_{basis}}_{mol,i,j}$, where $mol$ indexes molecules and $\matr{O}$ includes both $\matr{H}$ and $\matr{G}$.

At this point, we have transformed the aligned matrix elements input to the DFTB layer into the operator matrices needed for the Kohn-Sham equations of Eq.~\ref{eq:KSequations}.

We next compute the two-electron contributions to the Kohn-Sham equations, $\matr{H}^{(2)}$ of Eq.~\ref{eq:KSequations}, with a tensor operation for each value of $N_{basis}$,
\begin{equation}\label{eq:H2tensor}
\footnotesize
    \matr{H^{(2)}}^{N_{basis}}_{mol,i,j} = \frac{1}{2} \matr{S}^{N_{basis}}_{mol,i,j}
            \sum_{k} \left(\matr{G}^{N_{basis}}_{mol,i,k} + \matr{G}^{N_{basis}}_{mol,j,k} \right) \matr{\Delta Q}_{mol,k}
\end{equation}
The charge fluctuations, $\matr{\Delta Q}_{mol,k}$, are initialized from the starting model parameters and updated to obtain self consistency, as discussed below.

The generalized eigenvalue problem of Eq.~\ref{eq:KSequations} is converted to a self-adjoint eigenvalue problem. The overlap matrices are first diagonalized
\begin{equation}\label{eq:Sev}
    \matr{S}^{N_{basis}}_{mol,i,j} = \sum_k
       \matr{U}^{N_{basis}}_{mol,i,k}  \matr{\Lambda}^{N_{basis}}_{mol,k}
       \matr{U}^{N_{basis}}_{mol,j,k}
\end{equation}
and the results are used to form,
\begin{equation}\label{eq:phiS}
    \matr{\Phi}^{N_{basis}}_{mol,i,j} =
       \matr{U}^{N_{basis}}_{mol,i,j}  \left( \matr{\Lambda}^{N_{basis}}_{mol,j} \right)^{-1}
\end{equation}
A Fock operator for the self-adjoint eigensystem is then formed,
\begin{equation}\label{eq:Ftensor}
\footnotesize
    \matr{F}^{N_{basis}}_{mol,i,j} = \sum_{k,l}
       \matr{\Phi}^{N_{basis}}_{mol,k,i}
       \left( \matr{H}^{N_{basis}}_{mol,k,l} + \matr{H^{(2)}}^{N_{basis}}_{mol,k,l} \right)
       \matr{\Phi}^{N_{basis}}_{mol,l,j}
\end{equation}
and diagonalized
\begin{equation}\label{eq:Fevtensor}
    \matr{F}^{N_{basis}}_{mol,i,j} = \sum_k
       \matr{C'}^{N_{basis}}_{mol,i,k}  \matr{\epsilon}^{N_{basis}}_{mol,k}
       \matr{C'}^{N_{basis}}_{mol,j,k} .
\end{equation}
The tensors $\matr{\epsilon}^{N_{basis}}_{mol,k}$ hold the orbital energies, $\epsilon$ of Eq.~\ref{eq:KSequations}, and the orbital expansion coefficients, $C$ of Eq.~\ref{eq:orbs}, are given by
\begin{equation}\label{eq:orbtrans}
    \matr{C}^{N_{basis}}_{mol,i,j} = \sum_k
       \matr{\Phi}^{N_{basis}}_{mol,i,k}
       \matr{C'}^{N_{basis}}_{mol,k,j}
\end{equation}
The density matrix is obtained by first masking out the unoccupied orbitals,
\begin{equation}\label{eq:Cmask}
    \matr{Cocc}^{N_{basis}}_{mol,i,j} =
       \matr{Mask}^{N_{basis}}_{mol,i,j}
       \matr{C}^{N_{basis}}_{mol,i,j}
\end{equation}
where $\matr{Mask}^{N_{basis}}_{mol,i,k}$ is 1 for occupied orbitals $j$ and 0 otherwise. The density matrices are then
\begin{equation}\label{eq:rhotensor}
    \matr{\rho}^{N_{basis}}_{mol,i,j} = 2 \sum_k
       \matr{Cocc}^{N_{basis}}_{mol,i,k}
       \matr{Cocc}^{N_{basis}}_{mol,j,k} .
\end{equation}
The mask of Eq.~\ref{eq:Cmask} allows tensor operations to be used despite molecules with the same $N_{basis}$ having potentially different numbers of occupied molecular orbitals. The charge fluctuations, per atomic orbital, are then obtained from
\begin{equation}\label{eq:dQtensor}
    \matr{\Delta Q}^{N_{basis}}_{mol,i} =
     - \sum_{k} \matr{S}^{N_{basis}}_{mol,i,k}
              \matr{\rho}^{N_{basis}}_{mol,i,k}
    - \matr{Q^{(0)}}^{N_{basis}}_{mol,i}
\end{equation}
where $\matr{Q^{(0)}}^{N_{basis}}$ holds the references charges, $\matr{Q}^{(0)}$ of Eq.~\ref{eq:dqatom}. The electronic energy is then,
\begin{flalign}\label{eq:Eelectensor}
    \matr{E_{elec}}^{N_{basis}}_{mol} =
    &\sum_{i,j} \matr{\rho}^{N_{basis}}_{mol,i,j}
               \matr{H}^{N_{basis}}_{mol,i,j} \\
    &+ {1 \over 2} \sum_{i,j}
      \matr{\Delta Q}^{N_{basis}}_{mol,i}
      \matr{G}^{N_{basis}}_{mol,i,j}
      \matr{\Delta Q}^{N_{basis}}_{mol,j}
\end{flalign}
The molecular dipoles are obtained from $\matr{\Delta Q}^{N_{basis}}$ and the cartesian positions of the atoms.

Because the overlap matrices are not altered during model training, $\matr{\Phi}^{N_{basis}}$ may be computed once at the start of training and used throughout. However, because the overlap matrices depend on molecular geometry, $\matr{\Phi}^{N_{basis}}$ must be computed separately for each minibatch.

Given the computational graph for a forward pass through the DFTB layer, Tensorflow computes the gradients needed to train the model parameters. If the molecular orbitals become degenerate, the gradients associated with the eigensystem diverge. Any values in the gradient evaluation that are returned as undefined are set to zero. In assigning molecular geometries of a given isomer to train and test sets, we also sort such that train molecules are less likely to encounter degeneracies (see Supporting Information). Due to this, undefined values for the gradients occur very rarely.

\subsection{Self Consistent Field}\label{sec:scf}

The two-electron portion of the Hamiltonian depends on the current estimates for the atomic charge fluctuations, $\matr{\Delta Q}$ of Eq.~\ref{eq:H2tensor}. In application of DFTB to a new molecule, these charges are obtained through an iterative procedure that locates a fixed point, in which the charges used to construct $\matr{H^{(2)}}$ in Eq.~\ref{eq:H2tensor} agree with those generated from the use of $\matr{H^{(2)}}$, {\it i.e.} those predicted by Eq.~\ref{eq:dQtensor}. In the DFTB layer of Sec.~\ref{sec:tensors}, the charge distributions used to construct $\matr{H^{(2)}}$ are fed into the computational graph. Updating of the charges is therefore handled outside of the Tensorflow computational graph.

\begin{figure}[htb]
\centering
\includegraphics[width=0.95\columnwidth]{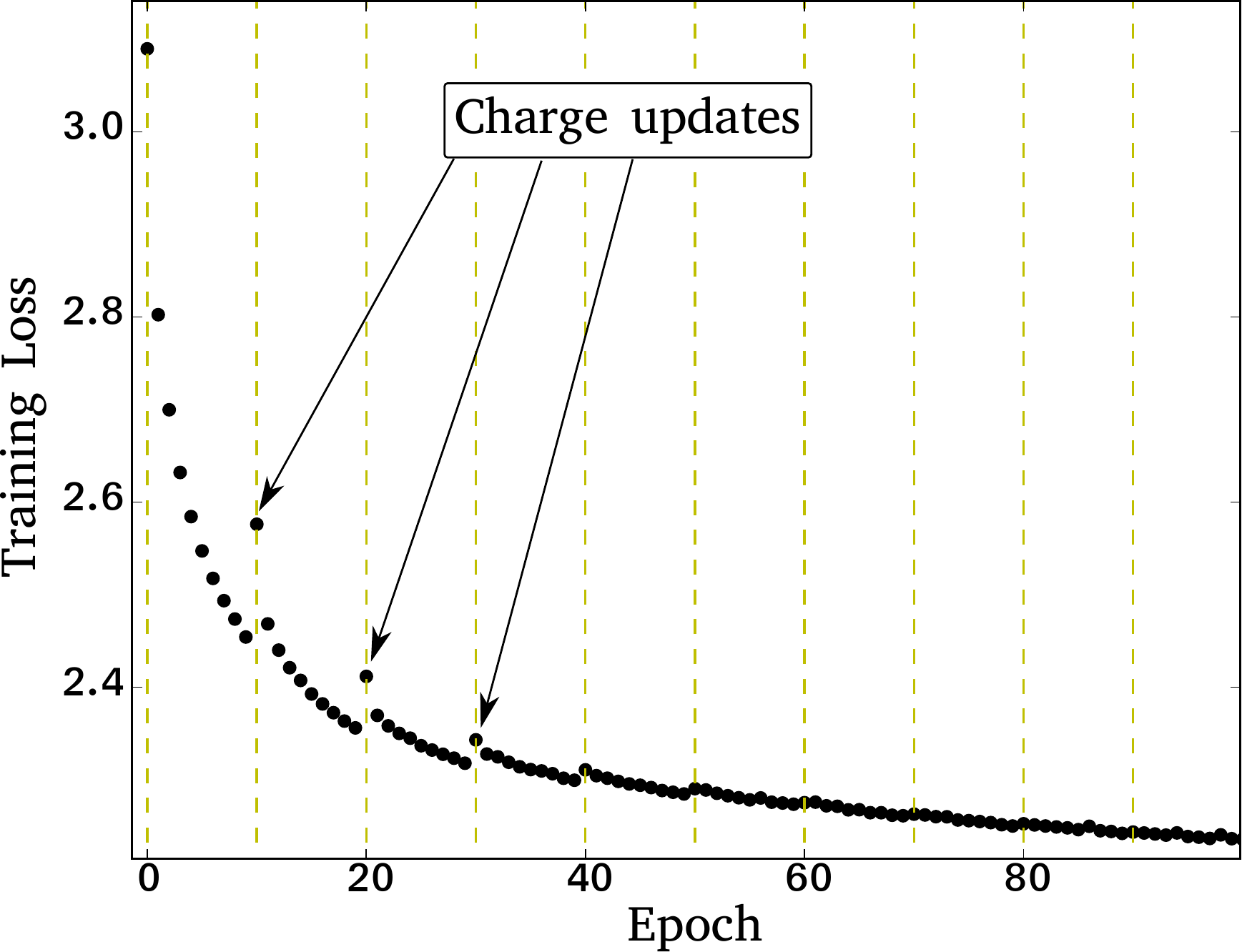}
\caption{Training error as a function of number of epochs. Every 10 epochs, the SCF solution for the current Hamiltonian parameters is used to update the charges fed into the DFTB layer. (Results are for a spline model with no regularization trained on molecules with up to 7 heavy atoms.) }
\label{fig:scfconverge}
\end{figure}

Roothaan-Hall is a simple iterative approach in which each iteration uses the charges generated by the previous iteration to update $\matr{H^{(2)}}$. Although such an approach is simple to implement, our experience is that this approach often fails to converge. We instead use the direct inverse of iterative subspace (DIIS) method\cite{pulay1982improved}. Periodically during training, the current $\matr{H}$ and $\matr{G}$ operator matrices are exported to a module that uses DIIS iterations to obtain the SCF charge distribution. The resulting charges are then fed into the computational graph, for use in constructing $\matr{H^{(2)}}$. Figure~\ref{fig:scfconverge} shows a representative training example in which the charges are updated every 10 epochs. Between charge updates, training brings the predictions closer into agreement with the target values, but this training ignores the impact that changes to the parameters have on the charges used to construct $\matr{H^{(2)}}$. When the charges are updated, the agreement between predicted and target values degrades as shown by the spike in the loss at 10 and 20 epochs in Figure~\ref{fig:scfconverge}. As training continues and the model parameters begin to stabilize, updates to charges have smaller impacts on the loss. For the remainder of this paper, we perform charge updates every 10 epochs and show only results obtained immediately following a charge update. These are the results that would be obtained if the current model parameters were used to obtain SCF solutions and so are an accurate reflection of model performance. Charge updates account for roughly half of the total computation time.

In our initial exploration of model forms, DIIS occasionally failed to converge. In such cases, we did not update the feed to the DFTB computational graph. The charges fed into the computational graph for such molecules were then those from the most recent converged DIIS procedure. While this approach allowed us to explore a wide variety of model forms, for the results presented here, DIIS converged in all cases.

\section{Dataset}\label{sec:dataset}

The results presented here are for molecules composed of the elements H, C and O from the ANI-1 dataset\cite{smith2017ani}. The molecular structures in this dataset were obtained by the Normal Mode Sampling (NMS) method\cite{smith2017ani} and reflect the types of structures that may arise in a room temperature simulation. Because the ANI-1 data set does not contain atomic charges, we used the GAUSSIAN program\cite{gaussian09} to generate data for each of the included molecules using the same quantum chemical method as used in ANI-1, i.e. Density Functional Theory with the $\omega$B97X functional and 6-31g(d) basis set. Electrostatic potential, ESP, charges were obtained using the technique of Hu et al, which gives charges that vary smoothly with changes in molecular geometry and basis set.\cite{hu2007fitting}

As discussed in Section~\ref{sec:tensors}, the computational graph depends on the empirical formulas of the molecules in each minibatch. We use a single computational graph corresponding to the 66 empirical formulas of Table~\ref{tab:minibatch}. The formulas are sorted by number of heavy atoms, with 10 molecules each for heavy atom counts of 1 to 5. For larger systems, the number of molecules in the minibatch drops roughly as the square of the number of heavy atoms. Because the number of aligned matrix elements in a molecule scales roughly quadratically with the number of heavy atoms, this choice distributes the aligned matrix elements computed for molecules with 5 or more heavy atoms roughly equally among different size molecules. The Supporting Information provides additional details on the selection of molecular geometries to include in the dataset.

\begin{table*}[t]
\centering
\caption{Data used to train the model consists of 200 train and 50 test minibatches. Each minibatch has 66 molecules with the indicated empirical formulas. For empirical formulas with greater than 250 isomers, indicated with a *, the isomers in the test set are distinct from those in the training set.}
\label{tab:minibatch}
\begin{tabular}{cp{3in}ccc}
\hline\hline
\multicolumn{1}{l}{\multirow{2}{*}{\begin{tabular}[c]{@{}l@{}}Number of \\ heavy atoms\end{tabular}}} & \multirow{2}{*}{Empirical Formulae}                            & \multicolumn{3}{c}{Number of molecules}      \\
\multicolumn{1}{l}{}                                                                                  &                                                               & minibatch   & train set    & test set      \\
\hline
1 & \ce{H2O} \ce{CH4} \ce{H_2O} \ce{CH_4} \ce{H_2O} \ce{CH_4} \ce{H_2O} \ce{CH_4} \ce{H_2O} \ce{CH_4}                       & 10          & 2000           & 500           \\
2 & \ce{H_2O_2} \ce{CH_4O} \ce{CH_2O} \ce{C_2H_2} \ce{C_2H_6} \ce{C_2H_4} \ce{H_2O_2} \ce{CH_4O} \ce{CH_2O} \ce{C_2H_2}             & 10          & 2000           & 500           \\
3 & \ce{C_2H_4O} \ce{C_2H_6O} \ce{C_3H_6} \ce{H_2O_3} \ce{CO_2} \ce{CH_2O_2} \ce{C_3H_4} \ce{C_3H_8} \ce{C_2H_4O} \ce{C_2H_6O}         & 10          & 2000           & 500           \\
4  & \ce{C_3H_6O} \ce{C_4H_8} \ce{C_2H_4O_2} \ce{C_3H_8O} \ce{C_3H_4O} \ce{C_4H_6} \ce{CH_2O_3} \ce{C_2H_2O_2} \ce{C_2H_6O_2} \ce{C_3H_2O}  & 10          & 2000           & 500           \\
5 & \ce{C_4H_8O} \ce{C_4H_6O} \ce{C_3H_6O_2} \ce{C_5H_{10}} \ce{C_3H_4O_2} \ce{C_5H_8} \ce{C_4H_{10}O} \ce{C_5H_6} \ce{C_3H_8O_2} \ce{C_4H_4O} & 10          & 2000           & 350           \\
6  & \ce{C_5H_{10}O} \ce{C_5H_8O} \ce{C_4H_8O_2} \ce{C_4H_6O_2} \ce{C_5H_6O} \ce{C_6H_{10}} \ce{C_6H_8}                   & 7           & 1400           & 350           \\
7  & \ce{C_6H_{10}O}* \ce{C_5H_8O_2}* \ce{C_6H_8O} \ce{C_6H_2O} \ce{C_5H_{10}O_2}                         & 5           & 1000           & 250           \\
8 & \ce{C_6H_{10}O_2}* \ce{C_7H_{10}O}* \ce{C_7H1_2O}* \ce{C_6H_8O_2}*                              & 4           &             & 200           \\
\multicolumn{1}{l}{}                                                                                  & \multicolumn{1}{r}{\textbf{Total:}}                           & \textbf{66} & \textbf{12400} & \textbf{3300} \\
\hline
\end{tabular}
\end{table*}
% \footnotesize{*Greater than 250 isomers, so test set has distinct set of isomers.}

% \footnotesize{**228 isomers, so test set has 28 distinct isomers, and 22 isomers that are also in train set.}

\section{Loss being minimized}
The model parameters are updated to minimize a weighted sum of the root mean square (RMS) error in the target molecular properties,
\begin{equation}\label{eq:loss}
    Loss = \sum_{prop} w_{prop} \sqrt{{1 \over N_{prop}} \sum_{i}^{N_{prop}}
    \left| Pred_i - Target_i \right|^2 }
\end{equation}
where $w_{prop}$ is the weight for property $prop$, $N_{prop}$ is the number of predicted quantities, $Pred_i$ is the value predicted from the current DFTB parameters and $Target_i$ is the target value from the DFT computations of Section~\ref{sec:dataset}.

The optimization begins with the DFTB parameters of Ref.~\cite{elstner1998self}, which we will refer to as mio-0-1.   The mio-0-1 parameter set is the first freely available DFTB parameter set developed for organic molecules including O, N, C, H.\cite{elstner1998self,kruger2005validation} The initial errors are shown in Table~\ref{tab:initialloss}. The error in energy depends on how the parameters for the reference energy of Eq.~\ref{eq:Eref} are obtained. In evaluating models below, we will fit either to molecules with up to 4 heavy atoms, or up to 7 heavy atoms. The initial errors for these are labeled {\it up to 4} and {\it up to 7} in Table~\ref{tab:initialloss}. In addition, the reference energy can be fit to either the total molecular energy, $E_{mol}$, or the energy per heavy atom, $E_{atom}$. The initial error in $E_{mol}$ increases substantially with molecular size while that in $E_{atom}$ is less dependent on size. In the fits shown below, we therefore use $E_{atom}$ in the loss function. It is also worth noting that, even when $E_{ref}$ is the only aspect of the model that is trained, performance for the energy is substantially enhanced by including molecules with up to 7 heavy atoms.

The target properties include the ESP charges and the cartesian components of the dipole moment computed from these ESP charges. Fitting to the ESP dipoles prevents competition between the dipole loss and the atomic charge loss. The DFTB Hamiltonian is also based on point charges and using the ESP dipole prevents the fitting process from considering aspects of the dipole moment that can not be accounted for within DFTB. Use of the ESP dipole has only minor effects, as indicated by the close agreement between initial error of the ESP dipole and actual dipole in Table~\ref{tab:initialloss}.

For all fits reported here, $w_{prop}$ is 1/(0.1 kcal/mol) for $E_{atom}$, 1/(0.01) for charges, and 1/(0.01 Debye) for dipole components. These values were chosen based on the relative magnitudes of the initial errors in Table~\ref{tab:initialloss}.

% \usepackage{multirow}
\begin{table*}[t]
\centering
\caption{RMS errors using initial DFTB parameters. $E_{mol}$ and $E_{atom}$ are from fitting the reference energy, $E_{ref}$ of Eq.~\protect\ref{eq:Eref}, to total energy versus energy per heavy atom. {\it up to 4} and {\it up to 7} are from fitting $E_{ref}$ to molecules with up to 4 versus up 7 heavy atoms. Dipole errors are for cartesian components of the dipole, with {\it ESP} indicating that the dipole is computed form the atomic charges. Units are kcal/mol for energy, $e^-$ for charges, and Debye for dipole.} \label{tab:initialloss}
\begin{tabular}{cccccclc}
\hline\hline
\multirow{2}{*}{\begin{tabular}[c]{@{}c@{}}Number of\\ heavy atoms\end{tabular}} & \multicolumn{2}{c}{$E_{mol}$} & \multicolumn{2}{c}{$E_{atom}$} & \multirow{2}{*}{\begin{tabular}[c]{@{}c@{}}Atomic\\ charges\end{tabular}} & \multirow{2}{*}{\begin{tabular}[c]{@{}l@{}} Actual \\ Dipole \end{tabular}} & \multirow{2}{*}{\begin{tabular}[c]{@{}c@{}}ESP \\ Dipole\end{tabular}} \\
 & up to 4 & up to 7 & up to 4 & up to 7 &  &  &  \\
\hline
1 & 3.51 & 4.16 & 1.95 & 2.30 & 0.20 & 0.228 & 0.244 \\
2 & 4.90 & 4.97 & 2.63 & 2.56 & 0.09 & 0.208 & 0.216 \\
3 & 6.99 & 7.12 & 2.57 & 2.62 & 0.14 & 0.154 & 0.152 \\
4 & 5.87 & 5.64 & 1.82 & 1.69 & 0.13 & 0.257 & 0.255 \\
5 & 7.86 & 7.38 & 1.99 & 1.67 & 0.15 & 0.233 & 0.233 \\
6 & 9.33 & 8.50 & 2.03 & 1.62 & 0.15 & 0.252 & 0.254 \\
7 & 12.22 & 10.79 & 2.18 & 1.71 & 0.17 & 0.313 & 0.312 \\
8 & 14.73 & 12.94 & 2.27 & 1.80 & 0.17 & 0.341 & 0.340 \\
\hline
\end{tabular}
\end{table*}

\section{Spline model}\label{sec:spline}

The DFTB layer of Sec.~\ref{sec:tensors} may be used to develop an semiempirical Hamiltonian in which the matrix elements depend only on the atomic elements and the distance between the atoms.  The resulting model is then directly compatible with current implements of DFTB.\cite{aradi2007dftb+}  In this approach, the diagonal matrix elements of Table~\ref{tab:models} depend only on the atomic element and orbital type while the off-diagonal elements add a dependence on interatomic distance, $r$.  The dependence on $r$ is handled through two cubic splines that meet at $r_c$, a cutoff distance that varies with element types. During training of the model, only the region below $r_c$ is varied.  Each of the two joined spline has 20 knots, with parameters initialized by a least-squares fit to the parameters of mio-0-1.\cite{elstner1998self} The boundary conditions at $r_c$ force a continuous zeroth and first derivative, with natural boundary conditions used at the outside extremes.

\begin{figure}[htb]
\centering
\includegraphics[width=0.95\columnwidth]{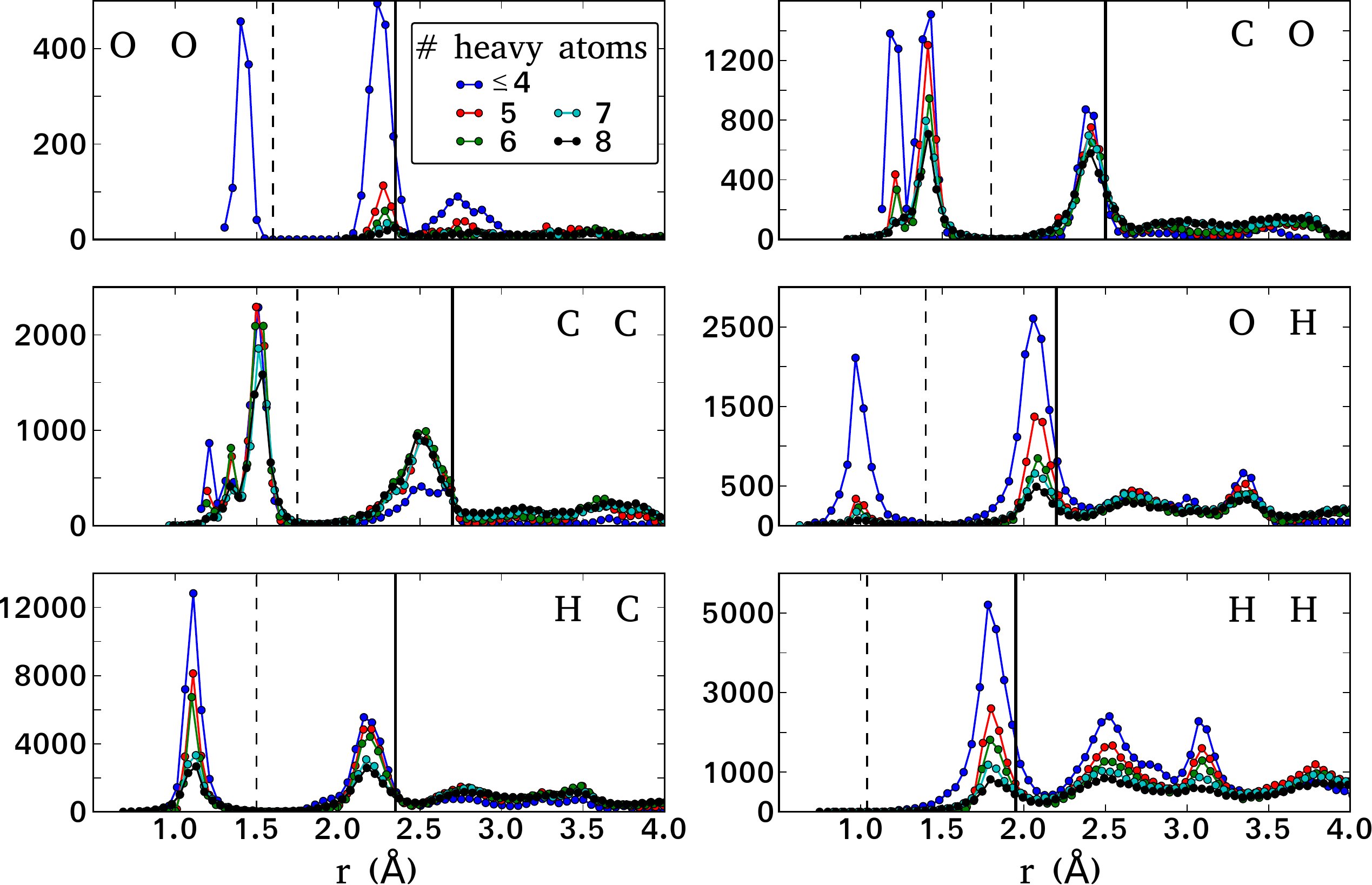}
\caption{Distribution of interatomic distances in the training and test data of Section.~\ref{sec:dataset}, for different pairs of elements. $x$-axis is inter-atomic distance in Angstroms and $y$-axis is number of occurrences in interval centered on that point. Vertical lines are cutoff distances for modifications to the repulsive potential, $R$, (dashed) and the electronic matrix elements, $H$ and $G$ (solid).}
\label{fig:bonddistances}
\end{figure}

%TODO (HC): why are the references for DFTB parameters here different than those for mio-0-1.
The values for $r_c$ were selected based on the distribution of interatomic distances in the dataset of Sec.~\ref{sec:dataset} (Fig.~\ref{fig:bonddistances}). For the repulsive potential, $R$, we use cutoffs just beyond the first peak in the distributions, such that the repulsive potential is varied only within distances corresponding to a covalent bond. This is consistent with the use of $R$ to model repulsive interactions between atomic cores. For the electronic matrix elements, $H$ and $G$, we choose $r_c$ just beyond the second peak in the distributions of Fig.~\ref{fig:bonddistances}. This allows the training of the model to modify next-nearest-neighbor electronic interactions between atoms. The model parameters are initialized to the matrix elements of mio-0-1.\cite{elstner1998self,niehaus2001application,sattelmeyer2006comparison} For $R$, the mio-0-1 matrix elements drop to near zero at $r_c$ and so the boundary conditions at $r_c$ constrain both the repulsive potential and its derivative to zero at $r_c$. For $H$ and $G$, the constraints at $r_c$ ensures a smooth transition to mio-0-1 values. For $G$, the mio-0-1 matrix elements tend towards Coulomb's law at long distances. The use of a constrained spline for $G$, initialized to mio-0-1 values beyond $r_c$, ensures that charge-charge interactions follow Coulomb's law at long distances. Transitioning from an empirical to physics-based model for charge-charge interactions may allow models trained on smaller molecules to transfer to larger molecules. The lower and upper ranges of the splines are set by the range of interatomic separations in Fig.~\ref{fig:bonddistances}.

The splines are implemented using a B-spline basis.\cite{de1978practical} Given the values at which the spline is to be evaluated, $r_i$, the values, $V_i$, can be obtained from a single matrix multiply,
\begin{equation}\label{eq:bspline}
    \matr{V} = \matr{B} \matr{X} + \matr{V}_0
\end{equation}
where $\matr{X}$ is a vector holding the parameters that may be varied to span the space of all cubic spline functions consistent with the above boundary conditions. $\matr{V}_0$ is a constant vector that is needed because the desired boundary conditions lead a general linear dependence on model parameters.
%$\matr{B}$ and $\matr{V}_0$ depend on the specific molecular geometries input to the computational graph, because these geometries establish the interatomic distances, $r_i$, at which the matrix elements must be evaluated. To account for this, $\matr{B}$ and $\matr{V}_0$ and initialized for each minibatch before the training begins and then included in the feed to the computational graph.

\begin{figure}[htb]
\centering
\includegraphics[width=0.95\columnwidth]{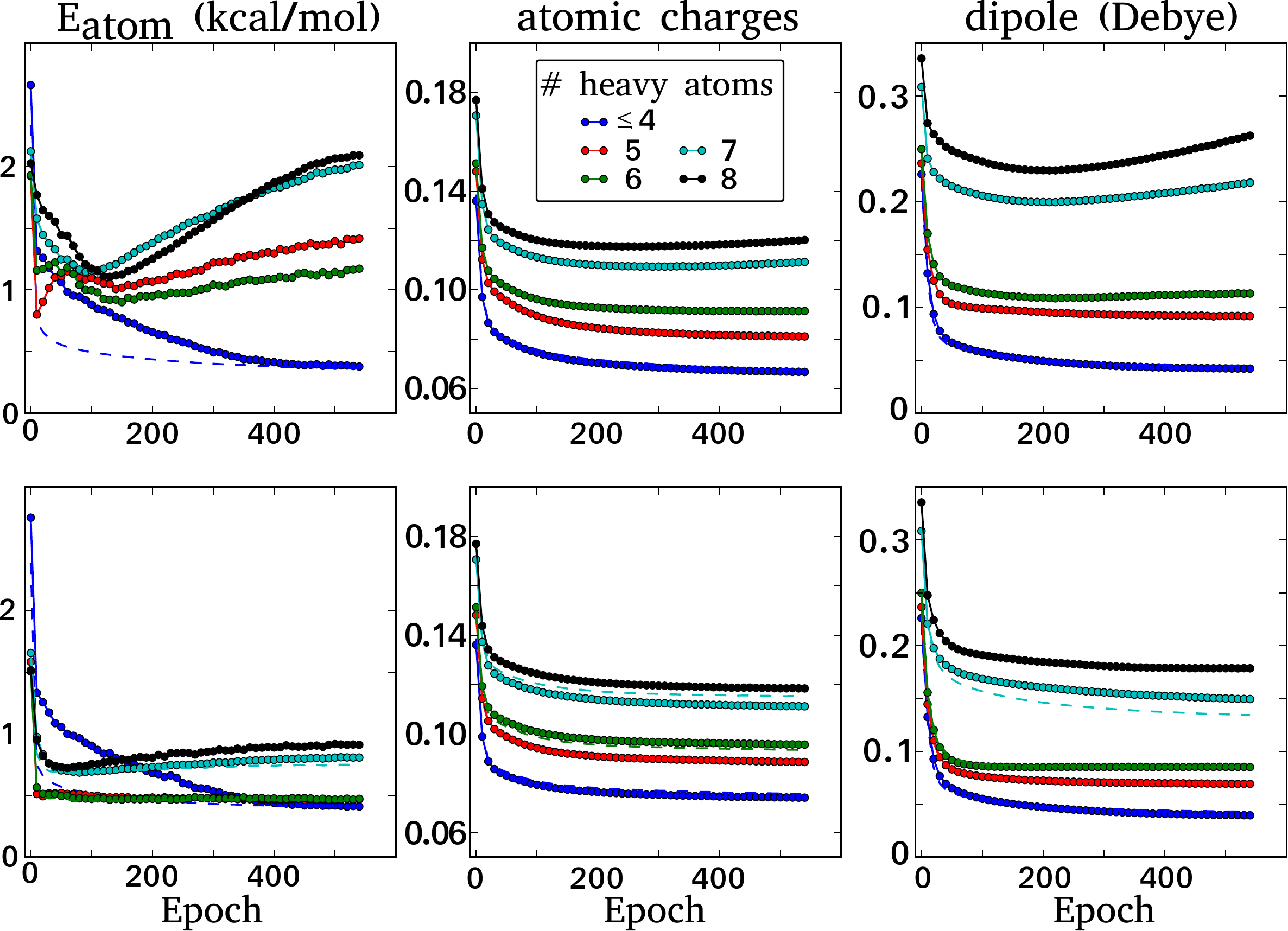}
\caption{Spline model without regularization. Upper and lower panels are trained on molecules with up to 4 and 7 heavy atoms, respectively. Solid lines are on test data and dotted lines are for train data. Charges are updated using DIIS to solve the self-consistent-field problem every 10 epochs, and data shown is following such updates.}
\label{fig:splinenoreg}
\end{figure}

\begin{figure}[htb]
\centering
\includegraphics[width=0.95\columnwidth]{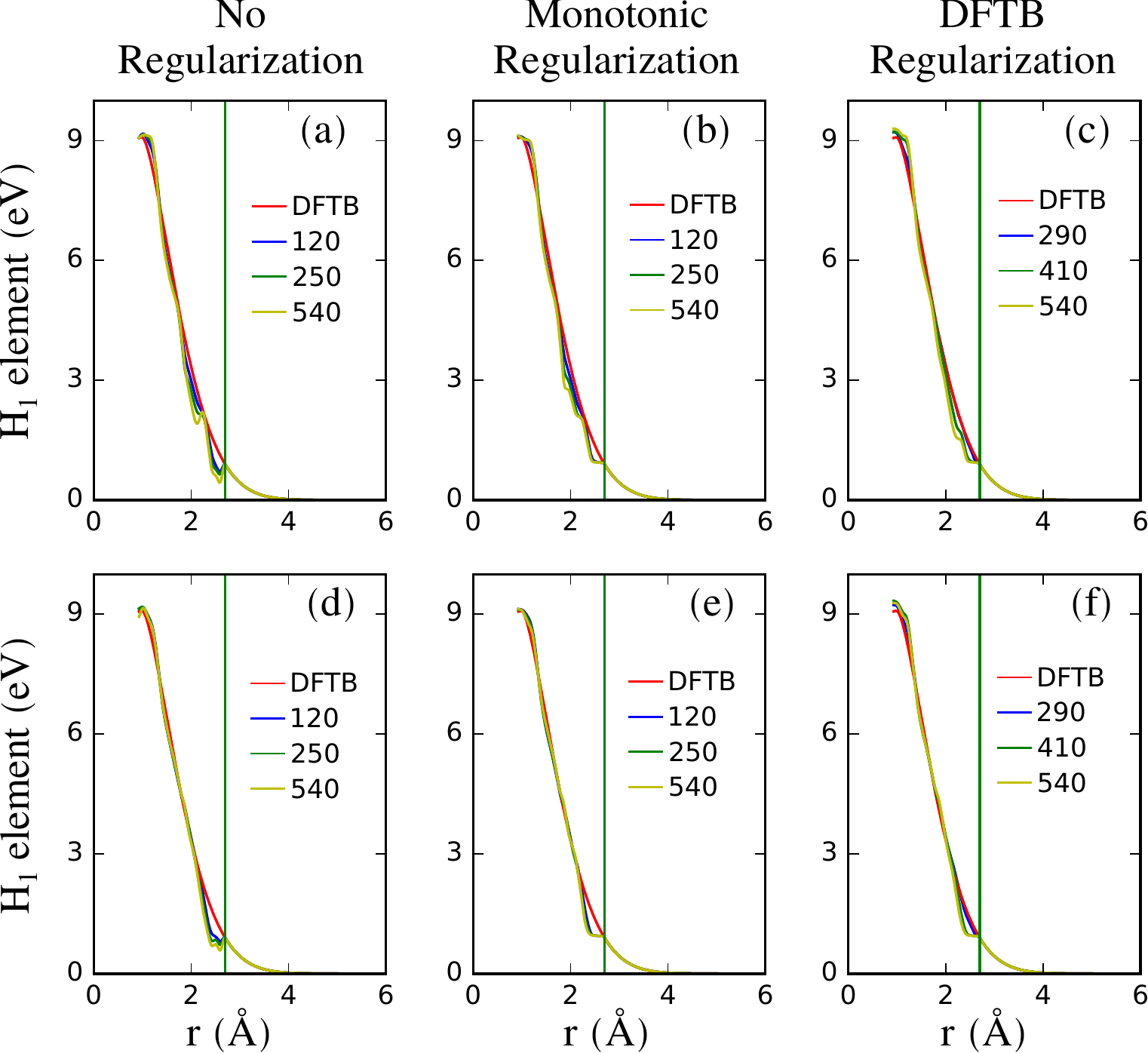}
\caption{Dependence on inter-atomic separation, $r$, of the one-electron Hamiltonian matrix element between $p$ orbitals of Carbon atoms, aligned in a $\sigma$ overlap orientation. Upper/lower panels are for models trained with up to 4 and up to 7 heavy atoms, respectively. The right-most column is with no regularization, the middle column is with monotonic regularization and the left-most column is for DFTB regularization. }
\label{fig:splinemodels}
\end{figure}

When the model is trained on molecules with up to 7 heavy atoms, the performance on test molecules with between 1 and 8 heavy atoms is improved, relative to the mio-0-1 values (lower panels of Fig.~\ref{fig:splinenoreg}). The RMS errors in energies and dipole components are reduced by about a factor of 2. However, when trained on molecules with up to 4 heavy atoms, the performance on molecules with more than 4 heavy atoms improves in early epochs but then begins to degrade (upper panels of Fig.~\ref{fig:splinenoreg}). Fig.~\ref{fig:splinemodels} shows how the spline evolves with training epoch, for matrix elements of $H$ between $p$ orbitals of carbon atoms aligned in a $\sigma$ orientation. Similar results are obtained for other matrix elements (see Supporting Information). When trained on molecules with up to 4 heavy atoms, the spline begins to oscillate (Fig.~\ref{fig:splinemodels}a) as the model begins to overtrain. When trained on molecules with up to 7 heavy atoms, the spline has a smoother dependence on $r$  (Fig.~\ref{fig:splinemodels}d). This suggests that overfitting may be due to overly complex forms for the dependence of the matrix elements on interatomic distance, $r$.

\begin{figure}[htb]
\centering
\includegraphics[width=0.95\columnwidth]{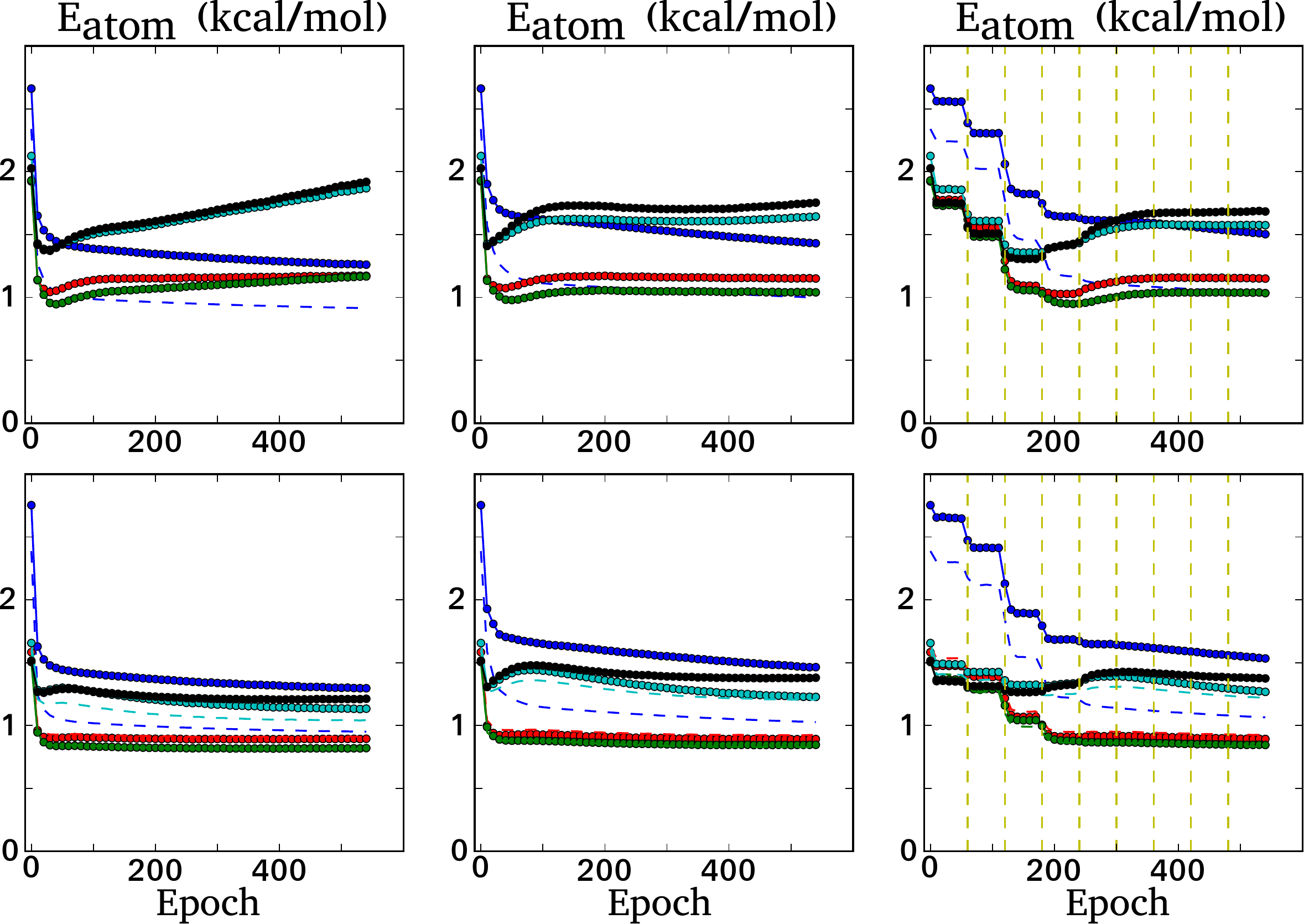}
\caption{Spline model used only for the repulsive energy. Upper and lower panels are for training on up to 4 and up to 7 heavy atoms, respectively. Left column uses no regularization and middle column uses monotonic regularization. Right column uses DFTB regularization, with dotted lines indicating points at which the penalty for deviation from mio-0-1 starting values is relaxed.}
\label{fig:splinereponly}
\end{figure}

\begin{figure}[htb]
\centering
\includegraphics[width=0.95\columnwidth]{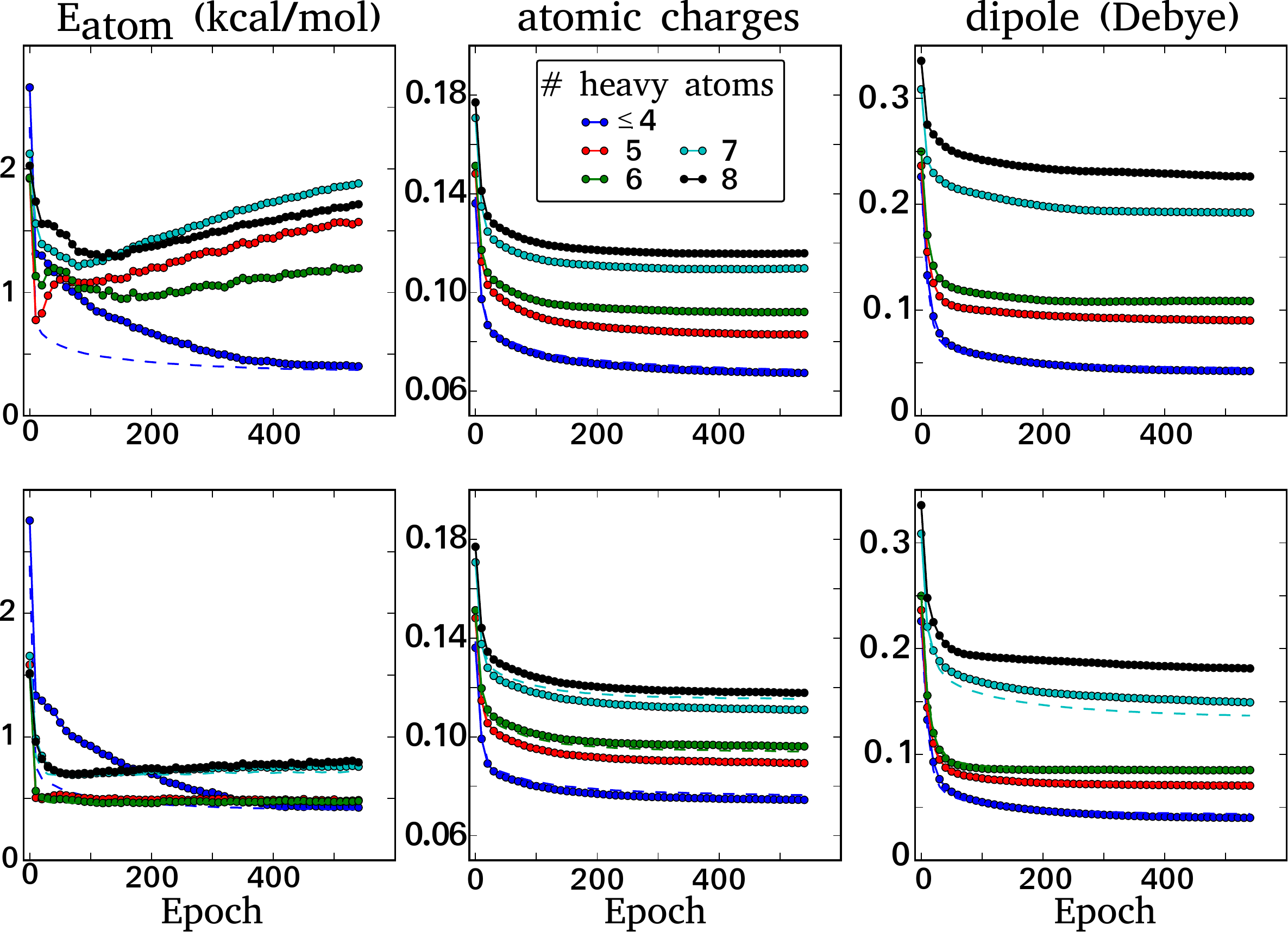}
\caption{Spline model with regularization used to force monotonic decay, using $\lambda_{mono}=10^5$ in Eq.~\ref{eq:monoreg}. Conventions are as in Fig.~\protect\ref{fig:splinenoreg}.}
\label{fig:splinemono}
\end{figure}

To help prevent overfitting, we tried imposing a penalty on oscillatory behaviors. This was done by adding an additional term to the loss of Eq.~\ref{eq:loss} that penalizes non-monotonic behavior,
\begin{equation}\label{eq:monoreg}
   Loss_{mono} = \lambda_{mono} \sum_i max(0,\,p V^\prime_i)^2
\end{equation}
where $\lambda_{mono}$ sets the magnitude of the penalty, the sum is over a uniform grid with the same range as the splines defining the matrix elements but with three times the density, $V^\prime_i$ is the derivative of the spline at the $i^{th}$ point of this dense grid, and $p$ is $+1$ if the matrix element should be a decreasing function of $r$ and $-1$ otherwise. This is implemented efficiently by using a linear form similar to that of Eq.~\ref{eq:bspline} to evaluate $V'_i$ and Tensorflow's $ReLU$ function for the $max$ term.  Values for $\lambda_{mono}$ of $10^3$ and $10^5$ give equivalent results, indicating that for both values, the penalty is sufficient to enforce monotonic behavior. Results shown in this paper are with $\lambda_{mono}$ = $10^5$. When only the repulsive potential $R$ is allowed to vary, monotonic regularization is sufficient to lead to improved transfer from smaller to larger molecules (panels $b$ and $e$ of Fig.~\ref{fig:splinereponly}). However, when the electronic Hamiltonian is also allowed to vary, monotonic regularization is not sufficient to improve transfer to larger molecules (Fig.~\ref{fig:splinemono}).  Fig.~\ref{fig:splinemodels}b suggest that, even with monotonic regularization, the dependence on $r$ remains overly complex.

\begin{figure}[htb]
\centering
\includegraphics[width=0.95\columnwidth]{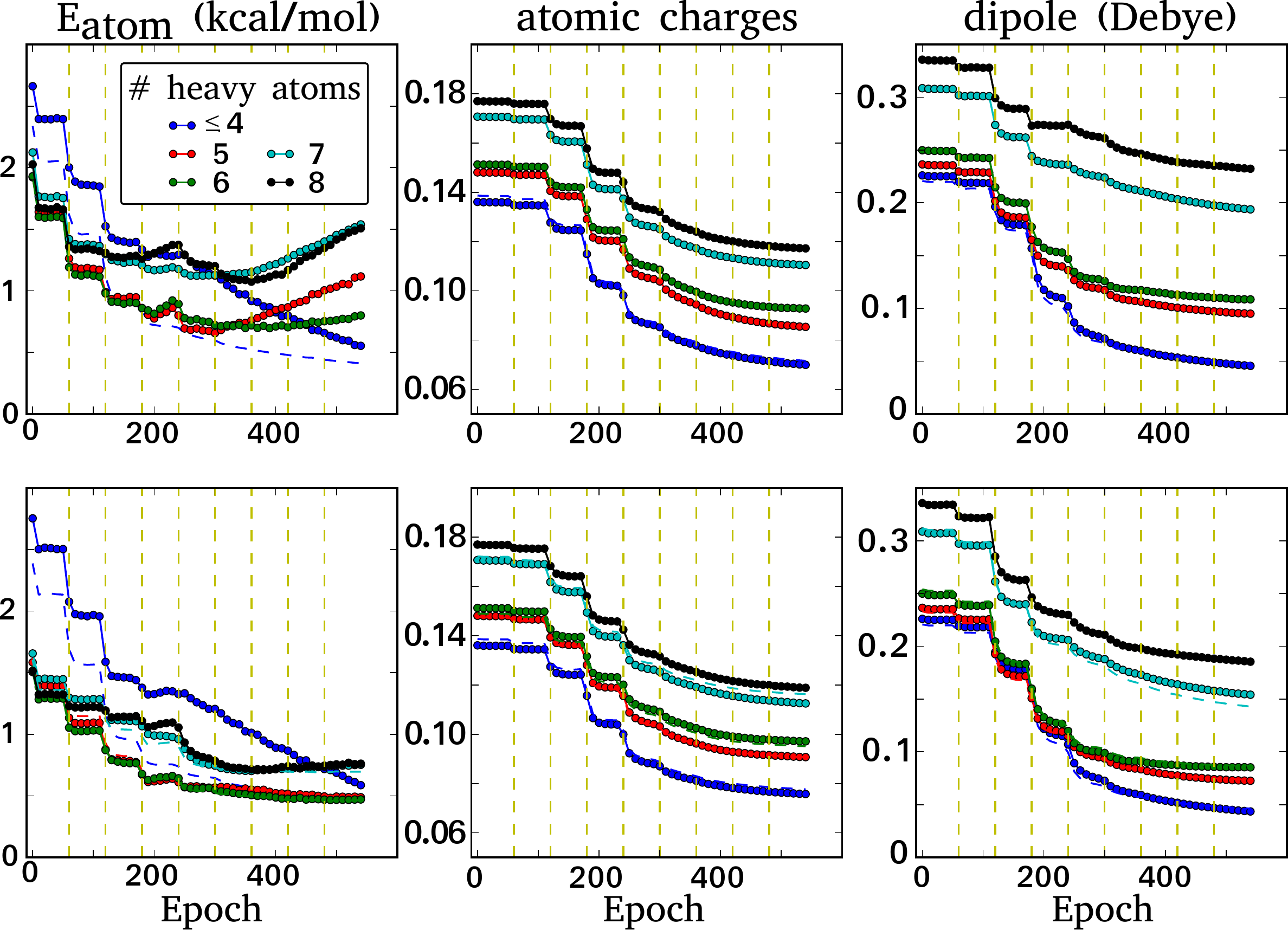}
\caption{Spline model with regularization used to force monotonic decay, using $\lambda_{mono}=10^5$ in Eq.~\ref{eq:monoreg}, and to penalize deviation from mio-0-1 values. Dotted lines indicate points at which the penalty for deviations from mio-0-1 starting values are relaxed, with values for $\lambda_{DFTB}$ following the series: 0.001, 0.003, 0.01, 0.03, 0.1, 0.3, 1, 3, and 10 kcal/mol. Remaining conventions are as in Fig.~\protect\ref{fig:splinenoreg}.}
\label{fig:splineres}
\end{figure}

As a means to further constrain the model space, we added an additional penalty for deviation from the original DFTB model.
\begin{equation}\label{eq:dftbreg}
\footnotesize
    Loss_{DFTB} = {1 \over \lambda_{DFTB}^2} {1 \over N_{elements}} \sum_i^{N_{elements}} (V_i - DFTB_i)^2
\end{equation}
where $\lambda_{DFTB}$ sets the magnitude of the penalty, the sum is over all matrix elements required for the minibatch, $V_i$ is the current value of the $i^{th}$ matrix element and $DFTB_i$ is the value of that element in the initial mio-0-1 parameterization.

Because the results have a strong dependence on $\lambda_{DFTB}$, for the remainder of this paper, we will present results in the format of Fig.~\ref{fig:splineres}. The DFTB penalty is relaxed during training by increasing the value of $\lambda_{DFTB}$ every 60 epochs. The vertical dotted lines in Fig.~\ref{fig:splineres} indicate epochs at which the penalty is relaxed. For values of $\lambda_{DFTB}$ where overfitting is not prevalent, 60 epochs is sufficient for training to stabilize. This approach therefore provides a concise picture of the dependence of the results on the DFTB penalty, before overfitting sets in, and the manner in which the model degrades, once overfitting sets in. The best performance is at about 360 epochs. Training on up to 7 heavy atoms and testing on 8 heavy atoms, leads to RMS errors in $E_{atom}$ of 0.72~kcal/mol, $q$ of 0.125, and $\mu$ of 0.20~D. Comparison with Table~\ref{tab:initialloss} shows that this corresponds to an error reduction of 60\% in $E_{atom}$, 27\% in $q$ and 41\% in $\mu$. Training on up to 4 heavy atoms gives similar improvements for $E_{atom}$ (1.08~kcal/mol, 53\%) and $q$ (0.12, 27\%), with less improvement seen for $\mu$ (0.25~D, 28\%). The values in parentheses are RMS errors and percent reduction relative to the column in Table~\ref{tab:initialloss} for which the reference energy was trained on molecules with up to 4 heavy atoms.

% (values are from using analyze.py to generate the plots and zooming in to extract the data from the graph.
% Trained on up to 7, tested on 8
%  mio-0-1 (from table 3):          1.80    0.17      0.340
%  Spline model at 360 epochs:    E 0.72  q 0.125  mu 0.198
%  percent as   (1.80-0.72)/0.80 : 0.6	0.2647058824	0.4176470588
% Trained on up to 4, tested on 8
%  mio-0-1:                      2.27     0.17      0.340
%  Spline model at 360 epochs: E 1.075  q 0.124  mu 0.246
%    percent:                    0.5264317181	0.2705882353	0.2764705882

The Supporting Information shows results from adjusting some classes of matrix elements during the fitting process,  freezing the remainder at their MIO-1-0 values. The results suggest that most of the improvement results from fitting the off-diagonal matrix elements of $\matr{H}$ and $\matr{G}$, and that simultaneous fitting of all matrix elements leads to the best performance.

%An additional advantage of the DFTB penalty is that it can be applied to any model form, including the neural networks of the next section where it is less apparent how to penalize deviations from monotonic behavior in $r$.

\section{Neural network models}\label{sec:nets}

\begin{figure}[htb]
\centering
\includegraphics[width=0.95\columnwidth]{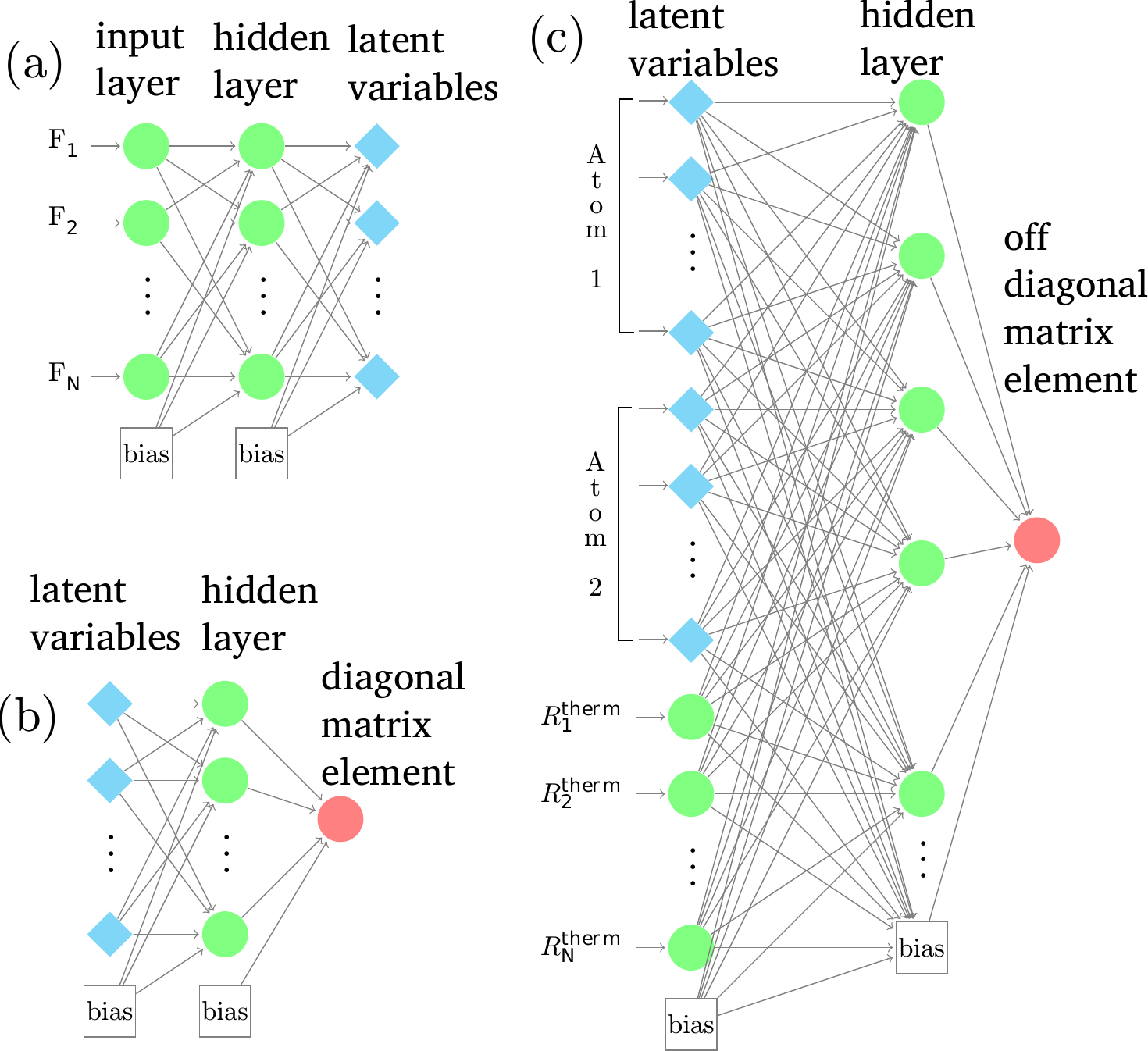}
\caption{Schematic representation of neural network structure, for the case of  1 hidden layer. (a) For each element type, C, H and O, a network takes in features that describe the atomic environment, $F_i$, and outputs a set of latent variables for that atom. (b) For each of the diagonal matrix element types in Table~\ref{tab:models}, a neural network predicts the matrix element from the latent variables of the associated atom. (c) Off-diagonal elements are predicted from the latent variables of the two associated atoms and a thermometer-encoded representation of the interatomic distance, Eq.~(\ref{eq:rtherm})}.
\label{fig:nnstruct}
\end{figure}

In the spline models of the previous section, the matrix elements depend only on the element type and the distance between atoms. Here, we use neural networks to allow the matrix elements to depend on the molecular environment of the atoms. The features input to the network are those developed for the ANI-1 neural network\cite{smith2017ani}. Encoded bonds features\cite{collins2018constant} were also tried but did not perform as well (see Supporting Information). The form of the network is shown schematically in Fig.~\ref{fig:nnstruct}. For each of the three element types (C, H and O), we create a FFNN that takes the features $F_i$ of a particular atom as input and generates latent variables that summarize the atomic environment of that atom (Fig.~\ref{fig:nnstruct}a). This is similar to the use of latent variables in the deep tensor neural network~\cite{schutt2017quantum}. Each atom gets a single set of latent variables that are used to predict all matrix elements associated with that atom.

For each of the diagonal matrix element types in Table~\ref{tab:models}, a FFNN is used to predict the matrix element from the latent variable of the associated atom (Fig.~\ref{fig:nnstruct}b).  For each of the off-diagonal matrix elements in Table~\ref{tab:models}, a FFNN is used to predict the matrix element from the latent variables of the two associated atoms, concatenated with features, $R^{therm}_i$, that thermometer encode the inter-atomic distance (Fig.~\ref{fig:nnstruct}c). The thermometer encoding is done via
%       res[i,j] = x/ (1 + x)
%                  with x = exp(-sigma * (r[i] - r_grid[j])
%          sigma = width/(r_grid[1] - r_grid[0])  width = 1
\begin{equation}
    R^{therm}_i = { e^{-\sigma (r - r^{grid}_i)} \over 1+  e^{-\sigma (r - r^{grid}_i)}}
\end{equation}\label{eq:rtherm}
where $r$ is the inter-atomic distance being encoded, $r^{grid}_i$ is a uniform grid of 10 points spanning the range of inter-atomic distances in the data set, and $\sigma$ is the inverse of the grid spacing. We emphasize that each atom has a single set of latent variables, that is used in predicting all matrix elements involving that atom.

\begin{figure}[htb]
\centering
\includegraphics[width=0.95\columnwidth]{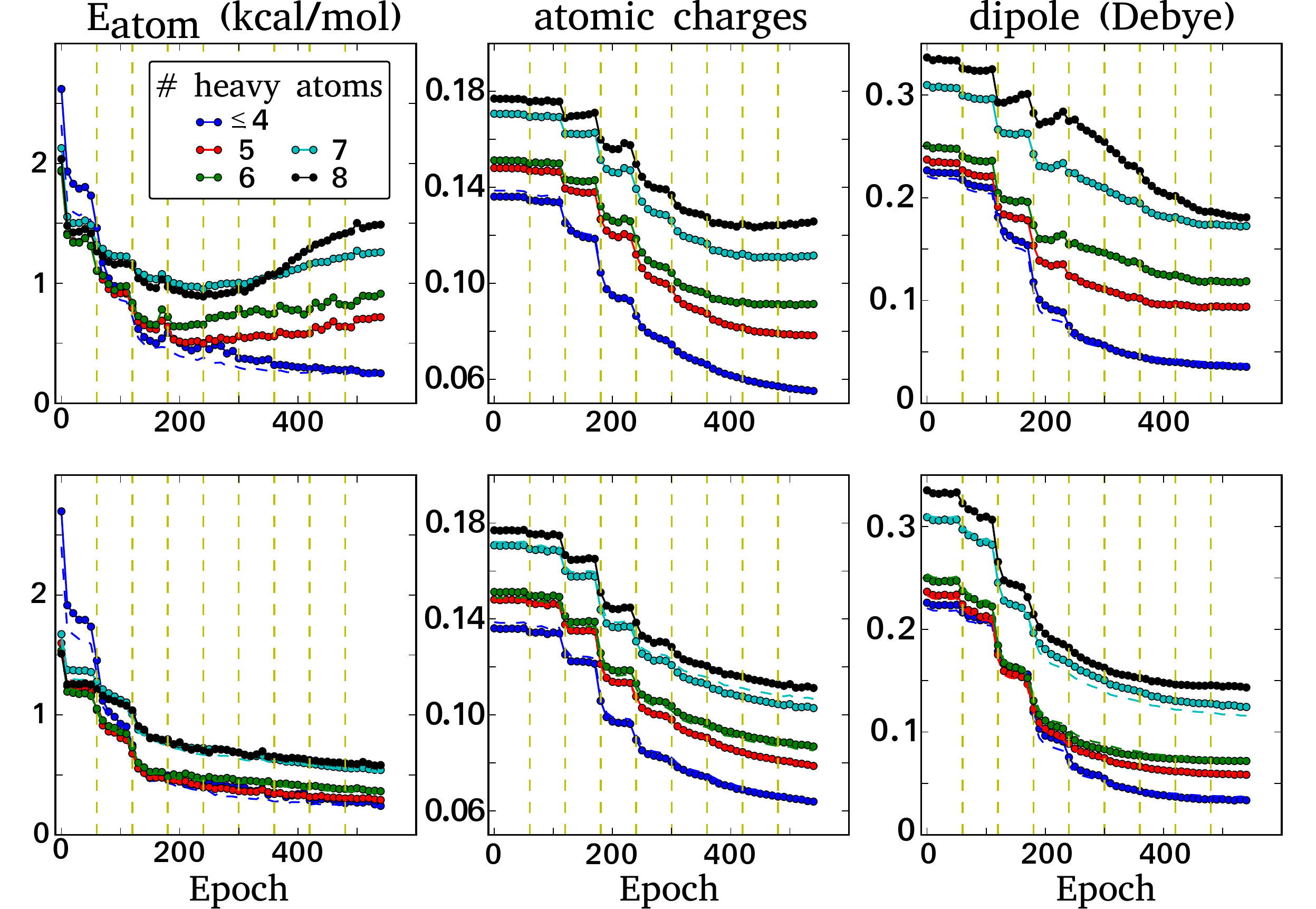}
\caption{Neural network model with regularization used to penalize deviations from mio-0-1 values. Conventions are as in Fig.~\protect\ref{fig:splineres}.}
\label{fig:nnres}
\end{figure}

Although the schematic of Fig.~\ref{fig:nnstruct} shows a single hidden layer for each of the networks, we have tried a number of other network structures (see Supporting Information). Fig.~\ref{fig:nnres} shows results from using ani1 features\cite{smith2017ani}, one hidden layer with 10 nodes for the latent network and the diagonal network, and one hidden layer with 50 nodes for the off-diagonal network. Sigmoid activation was used for all layers, except the final layer which is linear. Initialization occurs in two stages. In the first stage, the weights of each layer are initialized with a random normal distribution whose standard deviation is the inverse of the number of inputs to the layer. The distribution is truncated by redrawing any numbers with an absolute value greater than two standard deviations. In the second stage, the full network is then trained to reproduce the mio-0-1 matrix elements for all molecules in the training set for 5000 epochs, leading to an RMS error slightly below 0.1 kcal/mol. In all cases, the repulsive potential is treated with splines configured as in Sec.~\ref{sec:spline}.

% (values are from using analyze.py to generate the plots and zooming in to extract the data from the graph.
% Trained on up to 7, tested on 8
%  Spline model at 360 epochs:  E 0.72  q 0.125  mu 0.198
%  NN model model at 500 epochs:  0.59    0.113     0.145
%  percent as   (0.72-0.59)/0.72 : 0.1805555556	0.096	0.2676767677
% Trained on up to 4, tested on 8
%  Spline model at 360 epochs: E 1.075  q 0.124  mu 0.246
%  NN model at 360 epochs    : 1.06     0.128     0.225
%            percent         : 0.0139534884	-0.0322580645	0.0853658537
%  NN model at 240 epochs    : 0.89     0.150     0.275
%            percent         : 0.1720930233	-0.2096774194	-0.1178861789

The network is then trained using DFTB regularization in a manner identical to that used while training the spline models. Monotonic regularization is applied only to the repulsive potential. The results of Fig.~\ref{fig:nnres} show that the neural network model performs somewhat better than splines. Training on molecules with up to 7 heavy atoms and testing on molecules with 8 heavy atoms leads to RMS errors, at 500 epochs, in $E_{atom}$ of 0.59~kcal/mol, $q$ of 0.11, and $\mu$ of 0.14~D. This corresponds to an error reduction, relative to the spline model in Fig~\ref{fig:splineres}, of 18\% in $E_{atom}$, 10\% in $q$ and 27\% in $\mu$. When trained on molecules with up to 4 heavy atoms, the enhancement in performance on molecules with 8 heavy atoms, relative to the spline model, is more modest. At 240 epochs, the error in $E_{atom}$ is reduced by 18\% relative to the spline, but the errors in $q$ and $\mu$ are increased by 21\% and 12\% respectively. At 360 epochs, the error in all quantities is reduced relative to the spline model, but by only 1\% for $E_{atom}$, 3\% for $q$ and 9\% for $\mu$.

The Supporting Information includes results from a variety of other neural network architectures. These results show that making the model more flexible, by adding additional hidden layers or more nodes to the hidden layers, may somewhat improve performance when training on molecules with up to 7 heavy atoms, but degrades performance when training on molecules with up to 4 heavy atoms.  Additional approaches to regularization that better restrict the model flexibility may help with transfer from smaller to larger molecules, but developing such regularizations is left to further work.

\section{Discussion}

The DFTB layer for deep learning introduced here makes it computational feasible to adjust the electronic portion of a semiempirical QC Hamiltonian to relatively large sets of \textit{ab initio} data. Substantial improvements in predictions of energy, charge distributions and dipole moments were obtained with both the spline and neural network models explored here. Moreover, because the quantum chemical algorithm is implemented as a layer for deep leaning, any model that can be implemented in a deep learning framework may be used to generate the Hamiltonian matrix elements. This opens the possibility of discovering models that further improve performance and expand the range of chemical systems and properties that may be described.

A challenge in development of such models is that the resulting models are highly flexible. This flexibility comes from both the flexibility of models used to generate the Hamiltonian matrix elements and from the large number of such matrix elements present in each molecule. Detrimental effects from this flexibility were apparent in the overfitting that limited transfer of models trained on smaller molecules to larger models. This transfer was improved through regularization that penalized deviation of the matrix elements from those of the initial mio-0-1 model parameterization. Finding additional and more effective approaches to regularization is one avenue through which the utility of this approach can potentially be improved.

Despite the highly flexible nature of the model, performance on the training data remained above the 0.5~kcal/mol accuracy for total molecular energy that is the target for chemical accuracy. This indicates that the current model is not able to accurately capture the interactions present in the molecule. The similar performance observed for spline and neural network models suggests that the limitations are coming from the form of the Hamiltonian itself, as opposed to the models used to generate the matrix elements of this Hamiltonian.

Comparison of DFTB with other semiempirical models suggests one way in which the Hamiltonian can be generalized. In DFTB, the Coulomb interactions between atoms are described by charge-charge interactions. In Neglect of Diatomic Differential Overlap (NDDO) models, this is extended to include interactions between atomic dipoles and quadrupoles\cite{dewar1985development}. DFTB has also been extended to include dipole interactions.\cite{wu2017self} Extending the current quantum chemical layer to include higher multipole interactions should not lead, as far as we can anticipate, to any fundamental issues.

% TODO: Need to reference QUAMBO
Another source of information regarding more general forms of the model Hamiltonian comes from comparison with the results from quasiatomic minimal-basis orbitals (QUAMBO)\cite{quambo}. QUAMBO can, from the SCF solution of a molecule in a large basis set, generate a minimal basis Hamiltonian that will reproduce that SCF solution. The two-electron integrals of the resulting QUAMBO Hamiltonian are full four-index structures that do not adhere to the highly simplified forms of the DFTB or NDDO Hamiltonians. Inclusion of this aspect into the QC layer seems hopelessly complex. However, examination of QUAMBO Hamiltonians also reveals that the matrix elements between atomic $p$ orbitals do not exhibit the cylindrical symmetry implicit in the DFTB model form. This cylindrical symmetry arises from the use of a single $pp_{\pi}$ model (Table~\ref{tab:models}), to generate Hamiltonian matrix elements between the $p$ orbitals on two atoms. This assumption may be relaxed by allowing models to break the cylindrical symmetry, using features that describe the nonsymmetrical molecular environment about the relevant atoms.

In summary, we hope the DFTB layer developed here will aid development of parameterizations that expand the power and applicability of DFTB. In addition, we hope this will enable more systematic investigations into the benefits and challenges associated with incorporating quantum chemistry directly into deep learning models.

\section*{Acknowledgements}
The authors thank Adrian Roitberg for providing early access to the ANI-1 dataset. This work used the Extreme Science and Engineering Discovery Environment (XSEDE), which is supported by National Science Foundation grant number ACI-1548562. Specifically, it used the Bridges system, which is supported by NSF award number ACI-1445606, at the Pittsburgh Supercomputing Center (PSC). This work was also supported by the National Science Foundation under Grant CHE-1027985.

\clearpage

\bibliography{references}

\providecommand{\latin}[1]{#1}
\makeatletter
\providecommand{\doi}
  {\begingroup\let\do\@makeother\dospecials
  \catcode`\{=1 \catcode`\}=2\doi@aux}
\providecommand{\doi@aux}[1]{\endgroup\texttt{#1}}
\makeatother
\providecommand*\mcitethebibliography{\thebibliography}
\csname @ifundefined\endcsname{endmcitethebibliography}
  {\let\endmcitethebibliography\endthebibliography}{}
\begin{mcitethebibliography}{72}
\providecommand*\natexlab[1]{#1}
\providecommand*\mciteSetBstSublistMode[1]{}
\providecommand*\mciteSetBstMaxWidthForm[2]{}
\providecommand*\mciteBstWouldAddEndPuncttrue
  {\def\EndOfBibitem{\unskip.}}
\providecommand*\mciteBstWouldAddEndPunctfalse
  {\let\EndOfBibitem\relax}
\providecommand*\mciteSetBstMidEndSepPunct[3]{}
\providecommand*\mciteSetBstSublistLabelBeginEnd[3]{}
\providecommand*\EndOfBibitem{}
\mciteSetBstSublistMode{f}
\mciteSetBstMaxWidthForm{subitem}{(\alph{mcitesubitemcount})}
\mciteSetBstSublistLabelBeginEnd
  {\mcitemaxwidthsubitemform\space}
  {\relax}
  {\relax}

\bibitem[von Lilienfeld and Ramakrishnan(2017)von Lilienfeld, and
  Ramakrishnan]{von2017machine}
von Lilienfeld,~O.~A.; Ramakrishnan,~R. Machine learning, quantum chemistry,
  and chemical space. \emph{Reviews in computational chemistry} \textbf{2017},
  \emph{30}\relax
\mciteBstWouldAddEndPuncttrue
\mciteSetBstMidEndSepPunct{\mcitedefaultmidpunct}
{\mcitedefaultendpunct}{\mcitedefaultseppunct}\relax
\EndOfBibitem
\bibitem[Sanchez-Lengeling and Aspuru-Guzik(2018)Sanchez-Lengeling, and
  Aspuru-Guzik]{guzikInverse2018}
Sanchez-Lengeling,~B.; Aspuru-Guzik,~A. Inverse molecular design using machine
  learning: Generative models for matter engineering. \emph{Science}
  \textbf{2018}, \emph{361}, 360--365\relax
\mciteBstWouldAddEndPuncttrue
\mciteSetBstMidEndSepPunct{\mcitedefaultmidpunct}
{\mcitedefaultendpunct}{\mcitedefaultseppunct}\relax
\EndOfBibitem
\bibitem[Huan \latin{et~al.}(2015)Huan, Mannodi-Kanakkithodi, and
  Ramprasad]{rampi2015}
Huan,~T.~D.; Mannodi-Kanakkithodi,~A.; Ramprasad,~R. Accelerated Materials
  Property Predictions and Design Using Motif-based Fingerprints. \emph{Phys.
  Rev. B: Condens. Matter Mater. Phys.} \textbf{2015}, \emph{92}, 014106\relax
\mciteBstWouldAddEndPuncttrue
\mciteSetBstMidEndSepPunct{\mcitedefaultmidpunct}
{\mcitedefaultendpunct}{\mcitedefaultseppunct}\relax
\EndOfBibitem
\bibitem[Meredig \latin{et~al.}(2014)Meredig, Agrawal, Kirklin, Saal, Doak,
  Thompson, Zhang, Choudhary, and Wolverton]{ex_chem_screen2}
Meredig,~B.; Agrawal,~A.; Kirklin,~S.; Saal,~J.~E.; Doak,~J.; Thompson,~A.;
  Zhang,~K.; Choudhary,~A.; Wolverton,~C. Combinatorial Screening for New
  Materials in Unconstrained Composition Space with Machine Learning.
  \emph{Phys. Rev. B: Condens. Matter Mater. Phys.} \textbf{2014}, \emph{89},
  094104\relax
\mciteBstWouldAddEndPuncttrue
\mciteSetBstMidEndSepPunct{\mcitedefaultmidpunct}
{\mcitedefaultendpunct}{\mcitedefaultseppunct}\relax
\EndOfBibitem
\bibitem[O’Boyle \latin{et~al.}(2011)O’Boyle, Campbell, and
  Hutchison]{genetic}
O’Boyle,~N.~M.; Campbell,~C.~M.; Hutchison,~G.~R. Computational Design and
  Selection of Optimal Organic Photovoltaic Materials. \emph{J. Phys. Chem. C}
  \textbf{2011}, \emph{115}, 16200--16210\relax
\mciteBstWouldAddEndPuncttrue
\mciteSetBstMidEndSepPunct{\mcitedefaultmidpunct}
{\mcitedefaultendpunct}{\mcitedefaultseppunct}\relax
\EndOfBibitem
\bibitem[Rupp \latin{et~al.}(2012)Rupp, Tkatchenko, M{\"u}ller, and
  Von~Lilienfeld]{rupp2012fast}
Rupp,~M.; Tkatchenko,~A.; M{\"u}ller,~K.-R.; Von~Lilienfeld,~O.~A. Fast and
  accurate modeling of molecular atomization energies with machine learning.
  \emph{Physical review letters} \textbf{2012}, \emph{108}, 058301\relax
\mciteBstWouldAddEndPuncttrue
\mciteSetBstMidEndSepPunct{\mcitedefaultmidpunct}
{\mcitedefaultendpunct}{\mcitedefaultseppunct}\relax
\EndOfBibitem
\bibitem[Ruddigkeit \latin{et~al.}(2012)Ruddigkeit, van Deursen, Blum, and
  Reymond]{gdb17}
Ruddigkeit,~L.; van Deursen,~R.; Blum,~L.~C.; Reymond,~J.-L. Enumeration of 166
  Billion Organic Small Molecules in the Chemical Universe Database GDB-17.
  \emph{J. Chem. Inf. Comput. Sci.} \textbf{2012}, \emph{52}, 2864--2875\relax
\mciteBstWouldAddEndPuncttrue
\mciteSetBstMidEndSepPunct{\mcitedefaultmidpunct}
{\mcitedefaultendpunct}{\mcitedefaultseppunct}\relax
\EndOfBibitem
\bibitem[Hachmann \latin{et~al.}(2014)Hachmann, Olivares-Amaya, Jinich,
  Appleton, Blood-Forsythe, Seress, Rom\'{a}n-Salgado, Trepte, Atahan-Evrenk,
  Er, Shrestha, Mondal, Sokolov, Bao, and Aspuru-Guzik]{clean_energy2}
Hachmann,~J.; Olivares-Amaya,~R.; Jinich,~A.; Appleton,~A.~L.;
  Blood-Forsythe,~M.~A.; Seress,~L.~R.; Rom\'{a}n-Salgado,~C.; Trepte,~K.;
  Atahan-Evrenk,~S.; Er,~S.; Shrestha,~S.; Mondal,~R.; Sokolov,~A.; Bao,~Z.;
  Aspuru-Guzik,~A. Lead Candidates for High-performance Organic Photovoltaics
  from High-throughput Quantum Chemistry--The Harvard Clean Energy Project.
  \emph{Energy Environ. Sci.} \textbf{2014}, \emph{7}, 698--704\relax
\mciteBstWouldAddEndPuncttrue
\mciteSetBstMidEndSepPunct{\mcitedefaultmidpunct}
{\mcitedefaultendpunct}{\mcitedefaultseppunct}\relax
\EndOfBibitem
\bibitem[Hansen \latin{et~al.}(2013)Hansen, Montavon, Biegler, Fazli, Rupp,
  Scheffler, von Lilienfeld, Tkatchenko, and Müller]{atomization_review}
Hansen,~K.; Montavon,~G.; Biegler,~F.; Fazli,~S.; Rupp,~M.; Scheffler,~M.; von
  Lilienfeld,~O.~A.; Tkatchenko,~A.; Müller,~K.-R. Assessment and Validation
  of Machine Learning Methods for Predicting Molecular Atomization Energies.
  \emph{J. Chem. Theory Comput.} \textbf{2013}, \emph{9}, 3404--3419\relax
\mciteBstWouldAddEndPuncttrue
\mciteSetBstMidEndSepPunct{\mcitedefaultmidpunct}
{\mcitedefaultendpunct}{\mcitedefaultseppunct}\relax
\EndOfBibitem
\bibitem[Qu \latin{et~al.}(2013)Qu, Latino, and Aires-de Sousa]{split_energies}
Qu,~X.; Latino,~D.~A.; Aires-de Sousa,~J. A Big Data Approach to the Ultra-fast
  Prediction of DFT-calculated Bond Energies. \emph{J. Cheminf.} \textbf{2013},
  \emph{5}, 34\relax
\mciteBstWouldAddEndPuncttrue
\mciteSetBstMidEndSepPunct{\mcitedefaultmidpunct}
{\mcitedefaultendpunct}{\mcitedefaultseppunct}\relax
\EndOfBibitem
\bibitem[Sun \latin{et~al.}(2014)Sun, Wu, Song, Hu, Shan, and Chen]{nn_exp}
Sun,~J.; Wu,~J.; Song,~T.; Hu,~L.; Shan,~K.; Chen,~G. Alternative Approach to
  Chemical Accuracy: A Neural Networks-Based First-Principles Method for Heat
  of Formation of Molecules Made of H, C, N, O, F, S, and Cl. \emph{J. Phys.
  Chem. A} \textbf{2014}, \emph{118}, 9120--9131\relax
\mciteBstWouldAddEndPuncttrue
\mciteSetBstMidEndSepPunct{\mcitedefaultmidpunct}
{\mcitedefaultendpunct}{\mcitedefaultseppunct}\relax
\EndOfBibitem
\bibitem[Barker \latin{et~al.}(2017)Barker, Bulin, Hamaekers, and
  Mathias]{local_coulomb}
Barker,~J.; Bulin,~J.; Hamaekers,~J.; Mathias,~S. In \emph{Scientific Computing
  and Algorithms in Industrial Simulations: Projects and Products of Fraunhofer
  SCAI}; Griebel,~M., Sch{\"u}ller,~A., Schweitzer,~M.~A., Eds.; Springer
  International Publishing: Cham, 2017; pp 25--42\relax
\mciteBstWouldAddEndPuncttrue
\mciteSetBstMidEndSepPunct{\mcitedefaultmidpunct}
{\mcitedefaultendpunct}{\mcitedefaultseppunct}\relax
\EndOfBibitem
\bibitem[Yao \latin{et~al.}(2017)Yao, Herr, Brown, and
  Parkhill]{yao2017intrinsic}
Yao,~K.; Herr,~J.~E.; Brown,~S.~N.; Parkhill,~J. Intrinsic Bond Energies From a
  Bonds-in-Molecules Neural Network. \emph{J. Phys. Chem. Lett.} \textbf{2017},
  \emph{8}, 2689--2694\relax
\mciteBstWouldAddEndPuncttrue
\mciteSetBstMidEndSepPunct{\mcitedefaultmidpunct}
{\mcitedefaultendpunct}{\mcitedefaultseppunct}\relax
\EndOfBibitem
\bibitem[Ghiringhelli \latin{et~al.}(2015)Ghiringhelli, Vybiral, Levchenko,
  Draxl, and Scheffler]{systematic}
Ghiringhelli,~L.~M.; Vybiral,~J.; Levchenko,~S.~V.; Draxl,~C.; Scheffler,~M.
  Big Data of Materials Science: Critical Role of the Descriptor. \emph{Phys.
  Rev. Lett.} \textbf{2015}, \emph{114}, 105503\relax
\mciteBstWouldAddEndPuncttrue
\mciteSetBstMidEndSepPunct{\mcitedefaultmidpunct}
{\mcitedefaultendpunct}{\mcitedefaultseppunct}\relax
\EndOfBibitem
\bibitem[Collins \latin{et~al.}(2018)Collins, Gordon, von Lilienfeld, and
  Yaron]{collins2018constant}
Collins,~C.~R.; Gordon,~G.~J.; von Lilienfeld,~O.~A.; Yaron,~D.~J. Constant
  size descriptors for accurate machine learning models of molecular
  properties. \emph{The Journal of Chemical Physics} \textbf{2018}, \emph{148},
  241718\relax
\mciteBstWouldAddEndPuncttrue
\mciteSetBstMidEndSepPunct{\mcitedefaultmidpunct}
{\mcitedefaultendpunct}{\mcitedefaultseppunct}\relax
\EndOfBibitem
\bibitem[Wang \latin{et~al.}(2004)Wang, Wong, Hu, Chan, Su, and
  Chen]{early_delta}
Wang,~X.; Wong,~L.; Hu,~L.; Chan,~C.; Su,~Z.; Chen,~G. Improving the Accuracy
  of Density-Functional Theory Calculation: The Statistical Correction
  Approach. \emph{J. Phys. Chem. A} \textbf{2004}, \emph{108}, 8514--8525\relax
\mciteBstWouldAddEndPuncttrue
\mciteSetBstMidEndSepPunct{\mcitedefaultmidpunct}
{\mcitedefaultendpunct}{\mcitedefaultseppunct}\relax
\EndOfBibitem
\bibitem[Ediz \latin{et~al.}(2009)Ediz, Monda, Brown, and Yaron]{ediz_yaron}
Ediz,~V.; Monda,~A.~C.; Brown,~R.~P.; Yaron,~D.~J. Using Molecular Similarity
  to Develop Reliable Models of Chemical Reactions in Complex Environments.
  \emph{J. Chem. Theory Comput.} \textbf{2009}, \emph{5}, 3175--3184\relax
\mciteBstWouldAddEndPuncttrue
\mciteSetBstMidEndSepPunct{\mcitedefaultmidpunct}
{\mcitedefaultendpunct}{\mcitedefaultseppunct}\relax
\EndOfBibitem
\bibitem[Ramakrishnan \latin{et~al.}(2015)Ramakrishnan, Dral, Rupp, and von
  Lilienfeld]{ramakrishnan2015big}
Ramakrishnan,~R.; Dral,~P.~O.; Rupp,~M.; von Lilienfeld,~O.~A. Big data meets
  quantum chemistry approximations: the $\Delta$-machine learning approach.
  \emph{Journal of chemical theory and computation} \textbf{2015}, \emph{11},
  2087--2096\relax
\mciteBstWouldAddEndPuncttrue
\mciteSetBstMidEndSepPunct{\mcitedefaultmidpunct}
{\mcitedefaultendpunct}{\mcitedefaultseppunct}\relax
\EndOfBibitem
\bibitem[Ramakrishnan \latin{et~al.}(2015)Ramakrishnan, Hartmann, Tapavicza,
  and von Lilienfeld]{delta_ml_electronic}
Ramakrishnan,~R.; Hartmann,~M.; Tapavicza,~E.; von Lilienfeld,~O.~A. Electronic
  Spectra from TDDFT and Machine Learning in Chemical Space. \emph{J. Chem.
  Phys.} \textbf{2015}, \emph{143}, 084111\relax
\mciteBstWouldAddEndPuncttrue
\mciteSetBstMidEndSepPunct{\mcitedefaultmidpunct}
{\mcitedefaultendpunct}{\mcitedefaultseppunct}\relax
\EndOfBibitem
\bibitem[Smith \latin{et~al.}(2017)Smith, Isayev, and
  Roitberg]{smith2017anidata}
Smith,~J.~S.; Isayev,~O.; Roitberg,~A.~E. ANI-1, A data set of 20 million
  calculated off-equilibrium conformations for organic molecules.
  \emph{Scientific data} \textbf{2017}, \emph{4}, 170193\relax
\mciteBstWouldAddEndPuncttrue
\mciteSetBstMidEndSepPunct{\mcitedefaultmidpunct}
{\mcitedefaultendpunct}{\mcitedefaultseppunct}\relax
\EndOfBibitem
\bibitem[Smith \latin{et~al.}(2017)Smith, Isayev, and Roitberg]{smith2017ani}
Smith,~J.~S.; Isayev,~O.; Roitberg,~A.~E. ANI-1: an extensible neural network
  potential with DFT accuracy at force field computational cost. \emph{Chemical
  science} \textbf{2017}, \emph{8}, 3192--3203\relax
\mciteBstWouldAddEndPuncttrue
\mciteSetBstMidEndSepPunct{\mcitedefaultmidpunct}
{\mcitedefaultendpunct}{\mcitedefaultseppunct}\relax
\EndOfBibitem
\bibitem[Behler and Parrinello(2007)Behler, and
  Parrinello]{behler2007generalized}
Behler,~J.; Parrinello,~M. Generalized neural-network representation of
  high-dimensional potential-energy surfaces. \emph{Physical review letters}
  \textbf{2007}, \emph{98}, 146401\relax
\mciteBstWouldAddEndPuncttrue
\mciteSetBstMidEndSepPunct{\mcitedefaultmidpunct}
{\mcitedefaultendpunct}{\mcitedefaultseppunct}\relax
\EndOfBibitem
\bibitem[Behler()]{behler2015tutorial}
Behler,~J. Constructing high-dimensional neural network potentials: A tutorial
  review. \emph{International Journal of Quantum Chemistry} \emph{115},
  1032--1050\relax
\mciteBstWouldAddEndPuncttrue
\mciteSetBstMidEndSepPunct{\mcitedefaultmidpunct}
{\mcitedefaultendpunct}{\mcitedefaultseppunct}\relax
\EndOfBibitem
\bibitem[Boes \latin{et~al.}(2016)Boes, Groenenboom, Keith, and
  Kitchin]{boes2016neural}
Boes,~J.~R.; Groenenboom,~M.~C.; Keith,~J.~A.; Kitchin,~J.~R. Neural network
  and ReaxFF comparison for Au properties. \emph{International Journal of
  Quantum Chemistry} \textbf{2016}, \emph{116}, 979--987\relax
\mciteBstWouldAddEndPuncttrue
\mciteSetBstMidEndSepPunct{\mcitedefaultmidpunct}
{\mcitedefaultendpunct}{\mcitedefaultseppunct}\relax
\EndOfBibitem
\bibitem[Shen \latin{et~al.}(2016)Shen, Wu, and Yang]{shen2016multiscale}
Shen,~L.; Wu,~J.; Yang,~W. Multiscale quantum mechanics/molecular mechanics
  simulations with neural networks. \emph{Journal of chemical theory and
  computation} \textbf{2016}, \emph{12}, 4934--4946\relax
\mciteBstWouldAddEndPuncttrue
\mciteSetBstMidEndSepPunct{\mcitedefaultmidpunct}
{\mcitedefaultendpunct}{\mcitedefaultseppunct}\relax
\EndOfBibitem
\bibitem[Artrith and Urban(2016)Artrith, and Urban]{artrith2016implementation}
Artrith,~N.; Urban,~A. An implementation of artificial neural-network
  potentials for atomistic materials simulations: Performance for TiO2.
  \emph{Computational Materials Science} \textbf{2016}, \emph{114},
  135--150\relax
\mciteBstWouldAddEndPuncttrue
\mciteSetBstMidEndSepPunct{\mcitedefaultmidpunct}
{\mcitedefaultendpunct}{\mcitedefaultseppunct}\relax
\EndOfBibitem
\bibitem[Dewar \latin{et~al.}(1985)Dewar, Zoebisch, Healy, and
  Stewart]{dewar1985development}
Dewar,~M.~J.; Zoebisch,~E.~G.; Healy,~E.~F.; Stewart,~J.~J. Development and use
  of quantum mechanical molecular models. 76. AM1: a new general purpose
  quantum mechanical molecular model. \emph{Journal of the American Chemical
  Society} \textbf{1985}, \emph{107}, 3902--3909\relax
\mciteBstWouldAddEndPuncttrue
\mciteSetBstMidEndSepPunct{\mcitedefaultmidpunct}
{\mcitedefaultendpunct}{\mcitedefaultseppunct}\relax
\EndOfBibitem
\bibitem[Stewart(2013)]{stewart2013}
Stewart,~J. Optimization of parameters for semiempirical methods VI: More
  modifications to the NDDO approximations and re-optimization of parameters.
  \emph{J. Mol. Model.} \textbf{2013}, \emph{19}, 1--32\relax
\mciteBstWouldAddEndPuncttrue
\mciteSetBstMidEndSepPunct{\mcitedefaultmidpunct}
{\mcitedefaultendpunct}{\mcitedefaultseppunct}\relax
\EndOfBibitem
\bibitem[Thiel(2014)]{thiel2014semiempirical}
Thiel,~W. Semiempirical quantum--chemical methods. \emph{Wiley
  Interdisciplinary Reviews: Computational Molecular Science} \textbf{2014},
  \emph{4}, 145--157\relax
\mciteBstWouldAddEndPuncttrue
\mciteSetBstMidEndSepPunct{\mcitedefaultmidpunct}
{\mcitedefaultendpunct}{\mcitedefaultseppunct}\relax
\EndOfBibitem
\bibitem[Repasky \latin{et~al.}(2002)Repasky, Chandrasekhar, and
  Jorgensen]{repasky2002pddg}
Repasky,~M.~P.; Chandrasekhar,~J.; Jorgensen,~W.~L. PDDG/PM3 and PDDG/MNDO:
  improved semiempirical methods. \emph{Journal of computational chemistry}
  \textbf{2002}, \emph{23}, 1601--1622\relax
\mciteBstWouldAddEndPuncttrue
\mciteSetBstMidEndSepPunct{\mcitedefaultmidpunct}
{\mcitedefaultendpunct}{\mcitedefaultseppunct}\relax
\EndOfBibitem
\bibitem[Cui and Elstner(2014)Cui, and Elstner]{dftbReview2014}
Cui,~Q.; Elstner,~M. Density functional tight binding: values of semi-empirical
  methods in an ab initio era. \emph{Phys. Chem. Chem. Phys.} \textbf{2014},
  \emph{16}, 14368--14377\relax
\mciteBstWouldAddEndPuncttrue
\mciteSetBstMidEndSepPunct{\mcitedefaultmidpunct}
{\mcitedefaultendpunct}{\mcitedefaultseppunct}\relax
\EndOfBibitem
\bibitem[Girshick(2015)]{girshick2015fast}
Girshick,~R. Fast r-cnn. Proceedings of the IEEE international conference on
  computer vision. 2015; pp 1440--1448\relax
\mciteBstWouldAddEndPuncttrue
\mciteSetBstMidEndSepPunct{\mcitedefaultmidpunct}
{\mcitedefaultendpunct}{\mcitedefaultseppunct}\relax
\EndOfBibitem
\bibitem[Ren \latin{et~al.}(2015)Ren, He, Girshick, and Sun]{ren2015faster}
Ren,~S.; He,~K.; Girshick,~R.; Sun,~J. Faster r-cnn: Towards real-time object
  detection with region proposal networks. Advances in neural information
  processing systems. 2015; pp 91--99\relax
\mciteBstWouldAddEndPuncttrue
\mciteSetBstMidEndSepPunct{\mcitedefaultmidpunct}
{\mcitedefaultendpunct}{\mcitedefaultseppunct}\relax
\EndOfBibitem
\bibitem[Redmon \latin{et~al.}(2016)Redmon, Divvala, Girshick, and
  Farhadi]{redmon2016you}
Redmon,~J.; Divvala,~S.; Girshick,~R.; Farhadi,~A. You only look once: Unified,
  real-time object detection. Proceedings of the IEEE conference on computer
  vision and pattern recognition. 2016; pp 779--788\relax
\mciteBstWouldAddEndPuncttrue
\mciteSetBstMidEndSepPunct{\mcitedefaultmidpunct}
{\mcitedefaultendpunct}{\mcitedefaultseppunct}\relax
\EndOfBibitem
\bibitem[Collobert and Weston(2008)Collobert, and Weston]{collobert2008unified}
Collobert,~R.; Weston,~J. A unified architecture for natural language
  processing: Deep neural networks with multitask learning. Proceedings of the
  25th international conference on Machine learning. 2008; pp 160--167\relax
\mciteBstWouldAddEndPuncttrue
\mciteSetBstMidEndSepPunct{\mcitedefaultmidpunct}
{\mcitedefaultendpunct}{\mcitedefaultseppunct}\relax
\EndOfBibitem
\bibitem[Collobert \latin{et~al.}(2011)Collobert, Weston, Bottou, Karlen,
  Kavukcuoglu, and Kuksa]{collobert2011natural}
Collobert,~R.; Weston,~J.; Bottou,~L.; Karlen,~M.; Kavukcuoglu,~K.; Kuksa,~P.
  Natural language processing (almost) from scratch. \emph{Journal of Machine
  Learning Research} \textbf{2011}, \emph{12}, 2493--2537\relax
\mciteBstWouldAddEndPuncttrue
\mciteSetBstMidEndSepPunct{\mcitedefaultmidpunct}
{\mcitedefaultendpunct}{\mcitedefaultseppunct}\relax
\EndOfBibitem
\bibitem[Elstner \latin{et~al.}(1998)Elstner, Porezag, Jungnickel, Elsner,
  Haugk, Frauenheim, Suhai, and Seifert]{elstner1998self}
Elstner,~M.; Porezag,~D.; Jungnickel,~G.; Elsner,~J.; Haugk,~M.;
  Frauenheim,~T.; Suhai,~S.; Seifert,~G. Self-consistent-charge
  density-functional tight-binding method for simulations of complex materials
  properties. \emph{Physical Review B} \textbf{1998}, \emph{58}, 7260\relax
\mciteBstWouldAddEndPuncttrue
\mciteSetBstMidEndSepPunct{\mcitedefaultmidpunct}
{\mcitedefaultendpunct}{\mcitedefaultseppunct}\relax
\EndOfBibitem
\bibitem[Koskinen and M{\"a}kinen(2009)Koskinen, and
  M{\"a}kinen]{koskinen2009density}
Koskinen,~P.; M{\"a}kinen,~V. Density-functional tight-binding for beginners.
  \emph{Computational Materials Science} \textbf{2009}, \emph{47},
  237--253\relax
\mciteBstWouldAddEndPuncttrue
\mciteSetBstMidEndSepPunct{\mcitedefaultmidpunct}
{\mcitedefaultendpunct}{\mcitedefaultseppunct}\relax
\EndOfBibitem
\bibitem[Nishizawa \latin{et~al.}(2016)Nishizawa, Nishimura, Kobayashi, Irle,
  and Nakai]{nishizawa2016three}
Nishizawa,~H.; Nishimura,~Y.; Kobayashi,~M.; Irle,~S.; Nakai,~H. Three pillars
  for achieving quantum mechanical molecular dynamics simulations of huge
  systems: Divide-and-conquer, density-functional tight-binding, and massively
  parallel computation. \emph{Journal of computational chemistry}
  \textbf{2016}, \emph{37}, 1983--1992\relax
\mciteBstWouldAddEndPuncttrue
\mciteSetBstMidEndSepPunct{\mcitedefaultmidpunct}
{\mcitedefaultendpunct}{\mcitedefaultseppunct}\relax
\EndOfBibitem
\bibitem[Sattelmeyer \latin{et~al.}(2006)Sattelmeyer, Tirado-Rives, and
  Jorgensen]{sattelmeyer2006comparison}
Sattelmeyer,~K.~W.; Tirado-Rives,~J.; Jorgensen,~W.~L. Comparison of SCC-DFTB
  and NDDO-based semiempirical molecular orbital methods for organic molecules.
  \emph{The Journal of Physical Chemistry A} \textbf{2006}, \emph{110},
  13551--13559\relax
\mciteBstWouldAddEndPuncttrue
\mciteSetBstMidEndSepPunct{\mcitedefaultmidpunct}
{\mcitedefaultendpunct}{\mcitedefaultseppunct}\relax
\EndOfBibitem
\bibitem[Yang \latin{et~al.}(2008)Yang, Yu, York, Elstner, and
  Cui]{yang2008description}
Yang,~Y.; Yu,~H.; York,~D.; Elstner,~M.; Cui,~Q. Description of phosphate
  hydrolysis reactions with the self-consistent-charge
  density-functional-tight-binding (SCC-DFTB) theory. 1. Parameterization.
  \emph{Journal of chemical theory and computation} \textbf{2008}, \emph{4},
  2067--2084\relax
\mciteBstWouldAddEndPuncttrue
\mciteSetBstMidEndSepPunct{\mcitedefaultmidpunct}
{\mcitedefaultendpunct}{\mcitedefaultseppunct}\relax
\EndOfBibitem
\bibitem[Dolgonos \latin{et~al.}(2009)Dolgonos, Aradi, Moreira, and
  Frauenheim]{dolgonos2009improved}
Dolgonos,~G.; Aradi,~B.; Moreira,~N.~H.; Frauenheim,~T. An improved
  self-consistent-charge density-functional tight-binding (SCC-DFTB) set of
  parameters for simulation of bulk and molecular systems involving titanium.
  \emph{Journal of chemical theory and computation} \textbf{2009}, \emph{6},
  266--278\relax
\mciteBstWouldAddEndPuncttrue
\mciteSetBstMidEndSepPunct{\mcitedefaultmidpunct}
{\mcitedefaultendpunct}{\mcitedefaultseppunct}\relax
\EndOfBibitem
\bibitem[Wahiduzzaman \latin{et~al.}(2013)Wahiduzzaman, Oliveira, Philipsen,
  Zhechkov, van Lenthe, Witek, and Heine]{wahiduzzaman2013dftb}
Wahiduzzaman,~M.; Oliveira,~A.~F.; Philipsen,~P.; Zhechkov,~L.; van Lenthe,~E.;
  Witek,~H.~A.; Heine,~T. DFTB parameters for the periodic table: Part 1,
  Electronic structure. \emph{Journal of chemical theory and computation}
  \textbf{2013}, \emph{9}, 4006--4017\relax
\mciteBstWouldAddEndPuncttrue
\mciteSetBstMidEndSepPunct{\mcitedefaultmidpunct}
{\mcitedefaultendpunct}{\mcitedefaultseppunct}\relax
\EndOfBibitem
\bibitem[Oliveira \latin{et~al.}(2015)Oliveira, Philipsen, and
  Heine]{oliveira2015dftb}
Oliveira,~A.~F.; Philipsen,~P.; Heine,~T. DFTB parameters for the periodic
  table, part 2: energies and energy gradients from hydrogen to calcium.
  \emph{Journal of chemical theory and computation} \textbf{2015}, \emph{11},
  5209--5218\relax
\mciteBstWouldAddEndPuncttrue
\mciteSetBstMidEndSepPunct{\mcitedefaultmidpunct}
{\mcitedefaultendpunct}{\mcitedefaultseppunct}\relax
\EndOfBibitem
\bibitem[Zheng \latin{et~al.}(2007)Zheng, Witek, Bobadova-Parvanova, Irle,
  Musaev, Prabhakar, Morokuma, Lundberg, Elstner, K{\"o}hler, and
  Frauenheim]{zheng2007parameter}
Zheng,~G.; Witek,~H.~A.; Bobadova-Parvanova,~P.; Irle,~S.; Musaev,~D.~G.;
  Prabhakar,~R.; Morokuma,~K.; Lundberg,~M.; Elstner,~M.; K{\"o}hler,~C.;
  Frauenheim,~T. Parameter calibration of transition-metal elements for the
  spin-polarized self-consistent-charge density-functional tight-binding (DFTB)
  method: Sc, Ti, Fe, Co, and Ni. \emph{Journal of chemical theory and
  computation} \textbf{2007}, \emph{3}, 1349--1367\relax
\mciteBstWouldAddEndPuncttrue
\mciteSetBstMidEndSepPunct{\mcitedefaultmidpunct}
{\mcitedefaultendpunct}{\mcitedefaultseppunct}\relax
\EndOfBibitem
\bibitem[Knaup \latin{et~al.}(2007)Knaup, Hourahine, and
  Frauenheim]{knaup2007initial}
Knaup,~J.~M.; Hourahine,~B.; Frauenheim,~T. Initial Steps toward Automating the
  Fitting of DFTB Erep(r). \emph{The Journal of Physical Chemistry A}
  \textbf{2007}, \emph{111}, 5637--5641, PMID: 17428042\relax
\mciteBstWouldAddEndPuncttrue
\mciteSetBstMidEndSepPunct{\mcitedefaultmidpunct}
{\mcitedefaultendpunct}{\mcitedefaultseppunct}\relax
\EndOfBibitem
\bibitem[Gaus \latin{et~al.}(2009)Gaus, Chou, Witek, and
  Elstner]{gaus2009automatized}
Gaus,~M.; Chou,~C.-P.; Witek,~H.; Elstner,~M. Automatized parametrization of
  SCC-DFTB repulsive potentials: Application to hydrocarbons. \emph{The Journal
  of Physical Chemistry A} \textbf{2009}, \emph{113}, 11866--11881\relax
\mciteBstWouldAddEndPuncttrue
\mciteSetBstMidEndSepPunct{\mcitedefaultmidpunct}
{\mcitedefaultendpunct}{\mcitedefaultseppunct}\relax
\EndOfBibitem
\bibitem[Bodrog \latin{et~al.}(2011)Bodrog, Aradi, and
  Frauenheim]{bodrog2011automated}
Bodrog,~Z.; Aradi,~B.; Frauenheim,~T. Automated repulsive parametrization for
  the DFTB method. \emph{Journal of chemical theory and computation}
  \textbf{2011}, \emph{7}, 2654--2664\relax
\mciteBstWouldAddEndPuncttrue
\mciteSetBstMidEndSepPunct{\mcitedefaultmidpunct}
{\mcitedefaultendpunct}{\mcitedefaultseppunct}\relax
\EndOfBibitem
\bibitem[Louren{\c{c}}o \latin{et~al.}(2016)Louren{\c{c}}o, da~Silva, Oliveira,
  Quint{\~a}o, and Duarte]{lourencco2016fasp}
Louren{\c{c}}o,~M.~P.; da~Silva,~M.~C.; Oliveira,~A.~F.; Quint{\~a}o,~M.~C.;
  Duarte,~H.~A. FASP: a framework for automation of Slater--Koster file
  parameterization. \emph{Theoretical Chemistry Accounts} \textbf{2016},
  \emph{135}, 250\relax
\mciteBstWouldAddEndPuncttrue
\mciteSetBstMidEndSepPunct{\mcitedefaultmidpunct}
{\mcitedefaultendpunct}{\mcitedefaultseppunct}\relax
\EndOfBibitem
\bibitem[Abadi \latin{et~al.}(2016)Abadi, Barham, Chen, Chen, Davis, Dean,
  Devin, Ghemawat, Irving, Isard, Kudlur, Levenberg, Monga, Moore, Murray,
  Steiner, Tucker, Vasudevan, Warden, Wicke, Yu, and
  Zheng]{abadi2016tensorflow}
Abadi,~M.; Barham,~P.; Chen,~J.; Chen,~Z.; Davis,~A.; Dean,~J.; Devin,~M.;
  Ghemawat,~S.; Irving,~G.; Isard,~M.; Kudlur,~M.; Levenberg,~J.; Monga,~R.;
  Moore,~S.; Murray,~D.~G.; Steiner,~B.; Tucker,~P.; Vasudevan,~V.; Warden,~P.;
  Wicke,~M.; Yu,~Y.; Zheng,~X. Tensorflow: a system for large-scale machine
  learning. OSDI. 2016; pp 265--283\relax
\mciteBstWouldAddEndPuncttrue
\mciteSetBstMidEndSepPunct{\mcitedefaultmidpunct}
{\mcitedefaultendpunct}{\mcitedefaultseppunct}\relax
\EndOfBibitem
\bibitem[Nystrom \latin{et~al.}(2015)Nystrom, Levine, Roskies, and
  Scott]{bridges}
Nystrom,~N.~A.; Levine,~M.~J.; Roskies,~R.~Z.; Scott,~J.~R. Bridges: A Uniquely
  Flexible HPC Resource for New Communities and Data Analytics. Proceedings of
  the 2015 XSEDE Conference: Scientific Advancements Enabled by Enhanced
  Cyberinfrastructure. New York, NY, USA, 2015; pp 30:1--30:8\relax
\mciteBstWouldAddEndPuncttrue
\mciteSetBstMidEndSepPunct{\mcitedefaultmidpunct}
{\mcitedefaultendpunct}{\mcitedefaultseppunct}\relax
\EndOfBibitem
\bibitem[Kranz \latin{et~al.}(2018)Kranz, Kubillus, Ramakrishnan, von
  Lilienfeld, and Elstner]{kranz2018generalized}
Kranz,~J.~J.; Kubillus,~M.; Ramakrishnan,~R.; von Lilienfeld,~O.~A.;
  Elstner,~M. Generalized Density-Functional Tight-Binding Repulsive Potentials
  from Unsupervised Machine Learning. \emph{Journal of chemical theory and
  computation} \textbf{2018}, \emph{14}, 2341--2352\relax
\mciteBstWouldAddEndPuncttrue
\mciteSetBstMidEndSepPunct{\mcitedefaultmidpunct}
{\mcitedefaultendpunct}{\mcitedefaultseppunct}\relax
\EndOfBibitem
\bibitem[Chou \latin{et~al.}(2016)Chou, Nishimura, Fan, Mazur, Irle, and
  Witek]{chou2015automatized}
Chou,~C.-P.; Nishimura,~Y.; Fan,~C.-C.; Mazur,~G.; Irle,~S.; Witek,~H.~A.
  Automatized Parameterization of DFTB Using Particle Swarm Optimization.
  \emph{Journal of Chemical Theory and Computation} \textbf{2016}, \emph{12},
  53--64, PMID: 26587758\relax
\mciteBstWouldAddEndPuncttrue
\mciteSetBstMidEndSepPunct{\mcitedefaultmidpunct}
{\mcitedefaultendpunct}{\mcitedefaultseppunct}\relax
\EndOfBibitem
\bibitem[Krishnapriyan \latin{et~al.}(2017)Krishnapriyan, Yang, Niklasson, and
  Cawkwell]{cawkwell2017}
Krishnapriyan,~A.; Yang,~P.; Niklasson,~A. M.~N.; Cawkwell,~M.~J. Numerical
  Optimization of Density Functional Tight Binding Models: Application to
  Molecules Containing Carbon, Hydrogen, Nitrogen, and Oxygen. \emph{Journal of
  Chemical Theory and Computation} \textbf{2017}, \emph{13}, 6191--6200, PMID:
  29039935\relax
\mciteBstWouldAddEndPuncttrue
\mciteSetBstMidEndSepPunct{\mcitedefaultmidpunct}
{\mcitedefaultendpunct}{\mcitedefaultseppunct}\relax
\EndOfBibitem
\bibitem[Kullgren \latin{et~al.}(2017)Kullgren, Wolf, Hermansson, Köhler,
  Aradi, Frauenheim, and Broqvist]{balintDFTBauto2018}
Kullgren,~J.; Wolf,~M.~J.; Hermansson,~K.; Köhler,~C.; Aradi,~B.;
  Frauenheim,~T.; Broqvist,~P. Self-Consistent-Charge Density-Functional
  Tight-Binding (SCC-DFTB) Parameters for Ceria in 0D to 3D. \emph{The Journal
  of Physical Chemistry C} \textbf{2017}, \emph{121}, 4593--4607\relax
\mciteBstWouldAddEndPuncttrue
\mciteSetBstMidEndSepPunct{\mcitedefaultmidpunct}
{\mcitedefaultendpunct}{\mcitedefaultseppunct}\relax
\EndOfBibitem
\bibitem[Li and Qi(2018)Li, and Qi]{YueDFTBauto2018}
Li,~Y.; Qi,~Y. Transferable Self-Consistent Charge Density Functional
  Tight-Binding Parameters for Li–Metal and Li-Ions in Inorganic Compounds
  and Organic Solvents. \emph{The Journal of Physical Chemistry C}
  \textbf{2018}, \emph{122}, 10755--10764\relax
\mciteBstWouldAddEndPuncttrue
\mciteSetBstMidEndSepPunct{\mcitedefaultmidpunct}
{\mcitedefaultendpunct}{\mcitedefaultseppunct}\relax
\EndOfBibitem
\bibitem[Huran \latin{et~al.}(2018)Huran, Steigemann, Frauenheim, Aradi, and
  Marques]{huran2018efficient}
Huran,~A.~W.; Steigemann,~C.; Frauenheim,~T.; Aradi,~B.; Marques,~M. A.~L.
  Efficient Automatized Density-Functional Tight-Binding Parametrizations:
  Application to Group IV Elements. \emph{Journal of Chemical Theory and
  Computation} \textbf{2018}, \emph{14}, 2947--2954, PMID: 29733592\relax
\mciteBstWouldAddEndPuncttrue
\mciteSetBstMidEndSepPunct{\mcitedefaultmidpunct}
{\mcitedefaultendpunct}{\mcitedefaultseppunct}\relax
\EndOfBibitem
\bibitem[Weber and Thiel(2000)Weber, and Thiel]{weber2000orthogonalization}
Weber,~W.; Thiel,~W. Orthogonalization corrections for semiempirical methods.
  \emph{Theoretical Chemistry Accounts} \textbf{2000}, \emph{103},
  495--506\relax
\mciteBstWouldAddEndPuncttrue
\mciteSetBstMidEndSepPunct{\mcitedefaultmidpunct}
{\mcitedefaultendpunct}{\mcitedefaultseppunct}\relax
\EndOfBibitem
\bibitem[Dral \latin{et~al.}(2015)Dral, von Lilienfeld, and
  Thiel]{dral2015machine}
Dral,~P.~O.; von Lilienfeld,~O.~A.; Thiel,~W. Machine learning of parameters
  for accurate semiempirical quantum chemical calculations. \emph{Journal of
  chemical theory and computation} \textbf{2015}, \emph{11}, 2120--2125\relax
\mciteBstWouldAddEndPuncttrue
\mciteSetBstMidEndSepPunct{\mcitedefaultmidpunct}
{\mcitedefaultendpunct}{\mcitedefaultseppunct}\relax
\EndOfBibitem
\bibitem[Sch{\"u}tt and VandeVondele(2018)Sch{\"u}tt, and
  VandeVondele]{schutt2018machine}
Sch{\"u}tt,~O.; VandeVondele,~J. Machine learning adaptive basis sets for
  efficient large scale DFT simulation. \emph{Journal of chemical theory and
  computation} \textbf{2018}, \relax
\mciteBstWouldAddEndPunctfalse
\mciteSetBstMidEndSepPunct{\mcitedefaultmidpunct}
{}{\mcitedefaultseppunct}\relax
\EndOfBibitem
\bibitem[Kinga and Adam(2015)Kinga, and Adam]{kinga2015method}
Kinga,~D.; Adam,~J.~B. A method for stochastic optimization. International
  Conference on Learning Representations (ICLR). 2015\relax
\mciteBstWouldAddEndPuncttrue
\mciteSetBstMidEndSepPunct{\mcitedefaultmidpunct}
{\mcitedefaultendpunct}{\mcitedefaultseppunct}\relax
\EndOfBibitem
\bibitem[Pulay(1982)]{pulay1982improved}
Pulay,~P. \emph{J. Comput. Chem.} \textbf{1982}, \emph{3}, 556--560\relax
\mciteBstWouldAddEndPuncttrue
\mciteSetBstMidEndSepPunct{\mcitedefaultmidpunct}
{\mcitedefaultendpunct}{\mcitedefaultseppunct}\relax
\EndOfBibitem
\bibitem[Frisch \latin{et~al.}()Frisch, Trucks, Schlegel, Scuseria, Robb,
  Cheeseman, Scalmani, Barone, Mennucci, Petersson, Nakatsuji, Caricato, Li,
  Hratchian, Izmaylov, Bloino, Zheng, Sonnenberg, Hada, Ehara, Toyota, Fukuda,
  Hasegawa, Ishida, Nakajima, Honda, Kitao, Nakai, Vreven, Montgomery, Peralta,
  Ogliaro, Bearpark, Heyd, Brothers, Kudin, Staroverov, Kobayashi, Normand,
  Raghavachari, Rendell, Burant, Iyengar, Tomasi, Cossi, Rega, Millam, Klene,
  Knox, Cross, Bakken, Adamo, Jaramillo, Gomperts, Stratmann, Yazyev, Austin,
  Cammi, Pomelli, Ochterski, Martin, Morokuma, Zakrzewski, Voth, Salvador,
  Dannenberg, Dapprich, Daniels, Farkas, Foresman, Ortiz, Cioslowski, and
  Fox]{gaussian09}
Frisch,~M.~J.; Trucks,~G.~W.; Schlegel,~H.~B.; Scuseria,~G.~E.; Robb,~M.~A.;
  Cheeseman,~J.~R.; Scalmani,~G.; Barone,~V.; Mennucci,~B.; Petersson,~G.~A.;
  Nakatsuji,~H.; Caricato,~M.; Li,~X.; Hratchian,~H.~P.; Izmaylov,~A.~F.;
  Bloino,~J.; Zheng,~G.; Sonnenberg,~J.~L.; Hada,~M.; Ehara,~M.; Toyota,~K.;
  Fukuda,~R.; Hasegawa,~J.; Ishida,~M.; Nakajima,~T.; Honda,~Y.; Kitao,~O.;
  Nakai,~H.; Vreven,~T.; Montgomery,~J.~A.,~{Jr.}; Peralta,~J.~E.; Ogliaro,~F.;
  Bearpark,~M.; Heyd,~J.~J.; Brothers,~E.; Kudin,~K.~N.; Staroverov,~V.~N.;
  Kobayashi,~R.; Normand,~J.; Raghavachari,~K.; Rendell,~A.; Burant,~J.~C.;
  Iyengar,~S.~S.; Tomasi,~J.; Cossi,~M.; Rega,~N.; Millam,~J.~M.; Klene,~M.;
  Knox,~J.~E.; Cross,~J.~B.; Bakken,~V.; Adamo,~C.; Jaramillo,~J.;
  Gomperts,~R.; Stratmann,~R.~E.; Yazyev,~O.; Austin,~A.~J.; Cammi,~R.;
  Pomelli,~C.; Ochterski,~J.~W.; Martin,~R.~L.; Morokuma,~K.;
  Zakrzewski,~V.~G.; Voth,~G.~A.; Salvador,~P.; Dannenberg,~J.~J.;
  Dapprich,~S.; Daniels,~A.~D.; Farkas,~O.; Foresman,~J.~B.; Ortiz,~J.~V.;
  Cioslowski,~J.; Fox,~D.~J. \emph{Gaussian 09}; Revision D.01; Gaussian Inc.:
  Wallingford CT 2009\relax
\mciteBstWouldAddEndPuncttrue
\mciteSetBstMidEndSepPunct{\mcitedefaultmidpunct}
{\mcitedefaultendpunct}{\mcitedefaultseppunct}\relax
\EndOfBibitem
\bibitem[Hu \latin{et~al.}(2007)Hu, Lu, and Yang]{hu2007fitting}
Hu,~H.; Lu,~Z.; Yang,~W. Fitting molecular electrostatic potentials from
  quantum mechanical calculations. \emph{Journal of chemical theory and
  computation} \textbf{2007}, \emph{3}, 1004--1013\relax
\mciteBstWouldAddEndPuncttrue
\mciteSetBstMidEndSepPunct{\mcitedefaultmidpunct}
{\mcitedefaultendpunct}{\mcitedefaultseppunct}\relax
\EndOfBibitem
\bibitem[Kr{\"u}ger \latin{et~al.}(2005)Kr{\"u}ger, Elstner, Schiffels, and
  Frauenheim]{kruger2005validation}
Kr{\"u}ger,~T.; Elstner,~M.; Schiffels,~P.; Frauenheim,~T. Validation of the
  density-functional based tight-binding approximation method for the
  calculation of reaction energies and other data. \emph{The Journal of
  chemical physics} \textbf{2005}, \emph{122}, 114110\relax
\mciteBstWouldAddEndPuncttrue
\mciteSetBstMidEndSepPunct{\mcitedefaultmidpunct}
{\mcitedefaultendpunct}{\mcitedefaultseppunct}\relax
\EndOfBibitem
\bibitem[Aradi \latin{et~al.}(2007)Aradi, Hourahine, and
  Frauenheim]{aradi2007dftb+}
Aradi,~B.; Hourahine,~B.; Frauenheim,~T. DFTB+, a sparse matrix-based
  implementation of the DFTB method. \emph{The Journal of Physical Chemistry A}
  \textbf{2007}, \emph{111}, 5678--5684\relax
\mciteBstWouldAddEndPuncttrue
\mciteSetBstMidEndSepPunct{\mcitedefaultmidpunct}
{\mcitedefaultendpunct}{\mcitedefaultseppunct}\relax
\EndOfBibitem
\bibitem[Niehaus \latin{et~al.}(2001)Niehaus, Elstner, Frauenheim, and
  Suhai]{niehaus2001application}
Niehaus,~T.~A.; Elstner,~M.; Frauenheim,~T.; Suhai,~S. Application of an
  approximate density-functional method to sulfur containing compounds.
  \emph{Journal of Molecular Structure: THEOCHEM} \textbf{2001}, \emph{541},
  185--194\relax
\mciteBstWouldAddEndPuncttrue
\mciteSetBstMidEndSepPunct{\mcitedefaultmidpunct}
{\mcitedefaultendpunct}{\mcitedefaultseppunct}\relax
\EndOfBibitem
\bibitem[De~Boor \latin{et~al.}(1978)De~Boor, De~Boor, Math{\'e}maticien,
  De~Boor, and De~Boor]{de1978practical}
De~Boor,~C.; De~Boor,~C.; Math{\'e}maticien,~E.-U.; De~Boor,~C.; De~Boor,~C.
  \emph{A practical guide to splines}; Springer-Verlag New York, 1978;
  Vol.~27\relax
\mciteBstWouldAddEndPuncttrue
\mciteSetBstMidEndSepPunct{\mcitedefaultmidpunct}
{\mcitedefaultendpunct}{\mcitedefaultseppunct}\relax
\EndOfBibitem
\bibitem[Sch{\"u}tt \latin{et~al.}(2017)Sch{\"u}tt, Arbabzadah, Chmiela,
  M{\"u}ller, and Tkatchenko]{schutt2017quantum}
Sch{\"u}tt,~K.~T.; Arbabzadah,~F.; Chmiela,~S.; M{\"u}ller,~K.~R.;
  Tkatchenko,~A. Quantum-chemical insights from deep tensor neural networks.
  \emph{Nature communications} \textbf{2017}, \emph{8}, 13890\relax
\mciteBstWouldAddEndPuncttrue
\mciteSetBstMidEndSepPunct{\mcitedefaultmidpunct}
{\mcitedefaultendpunct}{\mcitedefaultseppunct}\relax
\EndOfBibitem
\bibitem[Wu \latin{et~al.}(2017)Wu, Ilie, and Crampin]{wu2017self}
Wu,~Y.; Ilie,~A.; Crampin,~S. Self-consistent charge and dipole density
  functional tight binding method and application to carbon-based systems.
  \emph{Computational Materials Science} \textbf{2017}, \emph{134},
  206--213\relax
\mciteBstWouldAddEndPuncttrue
\mciteSetBstMidEndSepPunct{\mcitedefaultmidpunct}
{\mcitedefaultendpunct}{\mcitedefaultseppunct}\relax
\EndOfBibitem
\bibitem[Lu \latin{et~al.}(2004)Lu, Wang, Schmidt, Bytautas, Ho, and
  Ruedenberg]{quambo}
Lu,~W.~C.; Wang,~C.~Z.; Schmidt,~M.~W.; Bytautas,~L.; Ho,~K.~M.; Ruedenberg,~K.
  Molecule intrinsic minimal basis sets. I. Exact resolution of ab initio
  optimized molecular orbitals in terms of deformed atomic minimal-basis
  orbitals. \emph{The Journal of Chemical Physics} \textbf{2004}, \emph{120},
  2629--2637\relax
\mciteBstWouldAddEndPuncttrue
\mciteSetBstMidEndSepPunct{\mcitedefaultmidpunct}
{\mcitedefaultendpunct}{\mcitedefaultseppunct}\relax
\EndOfBibitem
\end{mcitethebibliography}

\end{document}